\documentclass[10pt]{article}

\usepackage{authblk}

\usepackage{graphicx} 
\usepackage[scriptsize,nooneline,hang]{caption}
\usepackage[hang,nooneline,scriptsize]{subfigure}
\usepackage{pstricks,pstricks-add}
\usepackage{pst-plot}
\usepackage{color}

\usepackage{amsmath,amsfonts,bbm,dsfont,mathrsfs} 

\newcommand{\dd}{{\mathrm{d}}}
\newcommand{\tr}{\tilde{r}}
\newcommand{\tlt}{\tilde{t}}
\newcommand{\ta}{\tilde{a}}
\newcommand{\tL}{\tilde{L}}
\newcommand{\tR}{\tilde{R}}
\newcommand{\tXi}{\tilde{\Xi}}
\newcommand{\tA}{\tilde{A}}
\newcommand{\tB}{\tilde{B}}
\newcommand{\tTh}{\tilde{\Theta}}
\newcommand{\tK}{\tilde{K}}
\newcommand{\tN}{\tilde{N_g}}
\newcommand{\te}{\tilde{e}}
\newcommand{\tg}{\tilde{g}}
\newcommand{\tv}{\tilde{v}}

\begin{document}
\title{Analytic solutions of the geodesic equation for $\text{U(1)}^2$ dyonic rotating black holes} 

\author{Kai Flathmann}
\author{Saskia Grunau}
\affil{Institut f\"ur Physik, Universit\"at Oldenburg, D--26111 Oldenburg, Germany}

\maketitle

\begin{abstract}
 In this article we derive the geodesic equations in the $\text{U(1)}^2$ dyonic rotating black hole spacetime. We present their solutions in terms of the Kleinian $\sigma$-function and in special cases in terms of the Weierstra{\ss} $\wp$-, $\sigma$- and $\zeta$-functions. To give a list of all possible orbits, we analyse the geodesic motion of test particles and light using parametric diagrams and effective potentials. 
\end{abstract}

\section{Introduction}
Since many years Quantum Theory and General Relativity have passed all tests. Nevertheless the unification of both theories is still an open task. One interesting development in this case is the in late 1997 discovered AdS/CFT correspondence \cite{Maldacena:1998}. Therefore, a conformal field theory (CFT) acting on the boundary of an anti-de-Sitter (AdS) space is dual to a string theory with AdS background. Analyzing the structure of black holes in such a background could give insights on unsolved problems of the corresponding CFT. Another unsolved question is the problem of dark matter and dark energy, which possibly can be solved by introducing (pseudo-)scalar fields such as dilatons and axions \cite{Matarrese:2011,Guzman:1999}.   
One interesting spacetime containing both fields and a negative cosmological constant is the dyonic rotating black hole with four electromagnetic charges of the $\text{U(1)}^2$ gauged supergravity found by Chow and Comp\`{e}re \cite{Chow:2013gba}. This black hole has many subcases. With pairwise equal charges and vanishing NUT-charge, we obtain the Kerr-Newman-AdS black hole and the ungauged solution is similar to the Einstein-Maxwell-dilaton-axion (EMDA) black hole \cite{Garcia:1995qz}.

To understand the properties of a spacetime it is mandatory to study the geodesic behaviour of test particles and light and solve the equations of motion. In this article we derive these equations by separating the Hamilton-Jacobi equation, which separability can be shown by constructing Killing tensors \cite{Carter:1968,Walker:1970}. This has been done for a large class of black holes in \cite{Chervonyi:2015}. 

Previously analytical solutions were found in terms of the Weierstra{\ss} $\wp$-, $\sigma$- and $\zeta$-functions for the Schwarzschild-solution \cite{Hagihara:1931}, the Taub-NUT-solution \cite{Kagramanova:2010bk}, the Reissner-Nordstr\"om-solution \cite{Grunau:2010gd}, the five-dimensional Myers-Perry solution in the special case of equal rotation parameters \cite{Kagramanova:2012hw}, the Kerr-Newman spacetime for charged particles \cite{Hackmann:2013pva}, the Kerr-Newman-Taub-NUT spacetime \cite{Cebeci:2016} and for the EMDA spacetime \cite{Flathmann:2015}. In more general cases the solutions can be stated in terms of derivatives of the Kleinian $\sigma$-functions and were used to solve the geodesic equations in the four dimensional Schwarzschild-de Sitter \cite{Hackmann:2008zza, Hackmann:2008zz} and in various higher dimensional spacetimes like the Schwarzschild (-anti de Sitter) and Reissner-Nordstr\"om-solution \cite{Hackmann:2008tu}. The solutions were also found for the higher-dimensional Kerr (-anti de-Sitter) spacetime \cite{Hackmann:2010zz}, the higher dimensional Myers-Perry black hole \cite{Enolski:2010if}, the Ho\v{r}ava-Lifshitz black hole \cite{Enolski:2011id} and in $f(R)$-gravity \cite{Soroushfar:2015,Soroushfar:2016}.    

In this article we will derive the geodesic equations in the $\text{U(1)}^2$ dyonic rotating black hole spacetime and present their analytical solutions in two cases. First with vanishing cosmological constant or for light in terms of the Weierstra{\ss} $\wp$-, $\sigma$- and $\zeta$-functions and second in terms of the hyperelliptic Kleinian $\sigma$-functions. We will use parametric diagrams and effective potentials to analyze the radial and latitudinal motion and present a list of all possible orbit types afterwards. 
\section{The $\text{U(1)}^2$ dyonic rotating black hole}
For $\text{U(1)}^2$ dyonic rotating black holes in asymptotically AdS coordinates the metric is given by \cite{Chow:2013gba}
\begin{equation}
	\begin{split}
		\dd s^2 = &-\frac{R_g}{B-aA}\left(\dd t -\frac{A}{\Xi}\dd \phi\right)^2 + \frac{B-aA}{R_g}\dd r^2  \\
			  &+ \frac{\Theta_g a^2 \sin^2{\vartheta}}{B-aA}\left(\dd t-\frac{B}{a\Xi}\dd \phi\right)^2 + \frac{B-aA}{\Theta_g} \dd \vartheta^2 \, ,
	\end{split}
\end{equation}
with
\begin{equation}
 \begin{split}
  R_g &= r^2-2mr+a^2+e^2-N_g^2+g^2[r^4 \\
      &+(a^2+6N_g^2-2v^2)r^2+3N_g^2(a^2-N_g^2)]\\
  \Theta_g &= 1-a^2g^2\cos^2{\vartheta}-4a^2N_g \cos{\vartheta}\\
  A &= a\sin^2{\vartheta}+4N_g\sin^2{\frac{\vartheta}{2}}\\
  B &= r^2+(N_g+a)^2-v^2 \\ 
 \Xi &= 1-4N_g ag^2-a^2g^2 \, .
 \end{split}
\end{equation}
Here $m$ is the mass parameter, $a$ is the rotation parameter, $e$ and $v$ are charge dependant, $N_g$ corresponds to the Newman-Unti-Tamburino (NUT) charge and $g$ denotes the gauge coupling constant. For further information on the parameters see \cite{Chow:2013gba}. The coordinate system $(t, r, \vartheta, \phi)$ is Boyer-Lindquist like and the transformation to Cartesian like coordinates reads as
\begin{equation}
 	\begin{split}
		x&=\sqrt{(r^2+a^2)}\sin{\vartheta}\cos{\varphi} \\
		y&=\sqrt{(r^2+a^2)}\sin{\vartheta}\sin{\varphi} \\
		z&=r\cos{\vartheta} \, .
	\end{split}
\end{equation}

\begin{figure}[ht]
	\centering
	\includegraphics[width=0.3\textwidth]{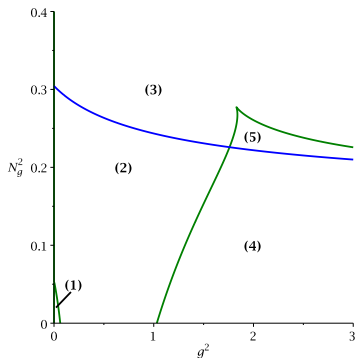}
	\caption{Parametric $g^2 - {N_g}^2$ - diagram for $a=0.4$, $e=0.38$, $v=1.3$ and $m=0.5$. The blue line represents $r=0$ as a zero and the green line shows where double zeros appear.}
	\label{pic:horizon-parametric-diagram}
\end{figure}

The horizon equation $R_g=0$ is a polynomial equation of order 4, whose solution can be stated analytically in the case of $g=0$ as
\begin{equation}
 r_{\pm} = m \pm \sqrt{m^2-a^2-e^2+N_g^2} \, ,
\end{equation}
which is similar to the Kerr-Newman-Taub-NUT black hole and also, in order to have real horizons, leads to the condition $m^2-a^2-e^2+N_g^2 \geq 0$ for the ungauged solution.\\
In the case of $g \neq 0$ the existence of four horizons, where at least one has to be negative, is possible. Here we identify the event-horizon, the cauchy-horizon and the cosmological horizon from known subcases. The negative valued horizon is called virtual. If we want to analyze the possible zeros of $R_g$ in general, we need to check, where double zeros appear. This leads to the condition
\begin{equation}
	R_g(r)=0 \quad \text{and} \quad \frac{\dd R_g (r)}{\dd r}=0 \, .
\end{equation}
The configuration of horizons also changes, if $r=0$ is a zero of $R_g$. Combining both leads to the parametric $g^2 - {N_g}^2$ - diagrams in Fig. \ref{pic:horizon-parametric-diagram}.

We distinguish between the following configurations
\begin{enumerate}
 \item Region (1): $R_g(r)$ has only complex zeros.
 \item Region (2): Two positive zeros.
 \item Region (3): One positive and one negative zero.
 \item Region (4): Two positive and two negative zeros.
 \item Region (5): One positive and three negative zeros. 
\end{enumerate}

The boundary of the ergoregion exists where $g_{tt}$ changes signs and therefore is defined by
\begin{equation}
 a^2\Theta_g\sin^2{\vartheta}-R_g=0\, ,
\end{equation}
which is similar to the Kerr-Newman-Taub-NUT black hole for $g=0$.

\begin{figure}[ht]
	\centering
	\subfigure[$N=0.2$, $a=0.3$ and $v=0.1$]{
		\includegraphics[width=0.3\textwidth]{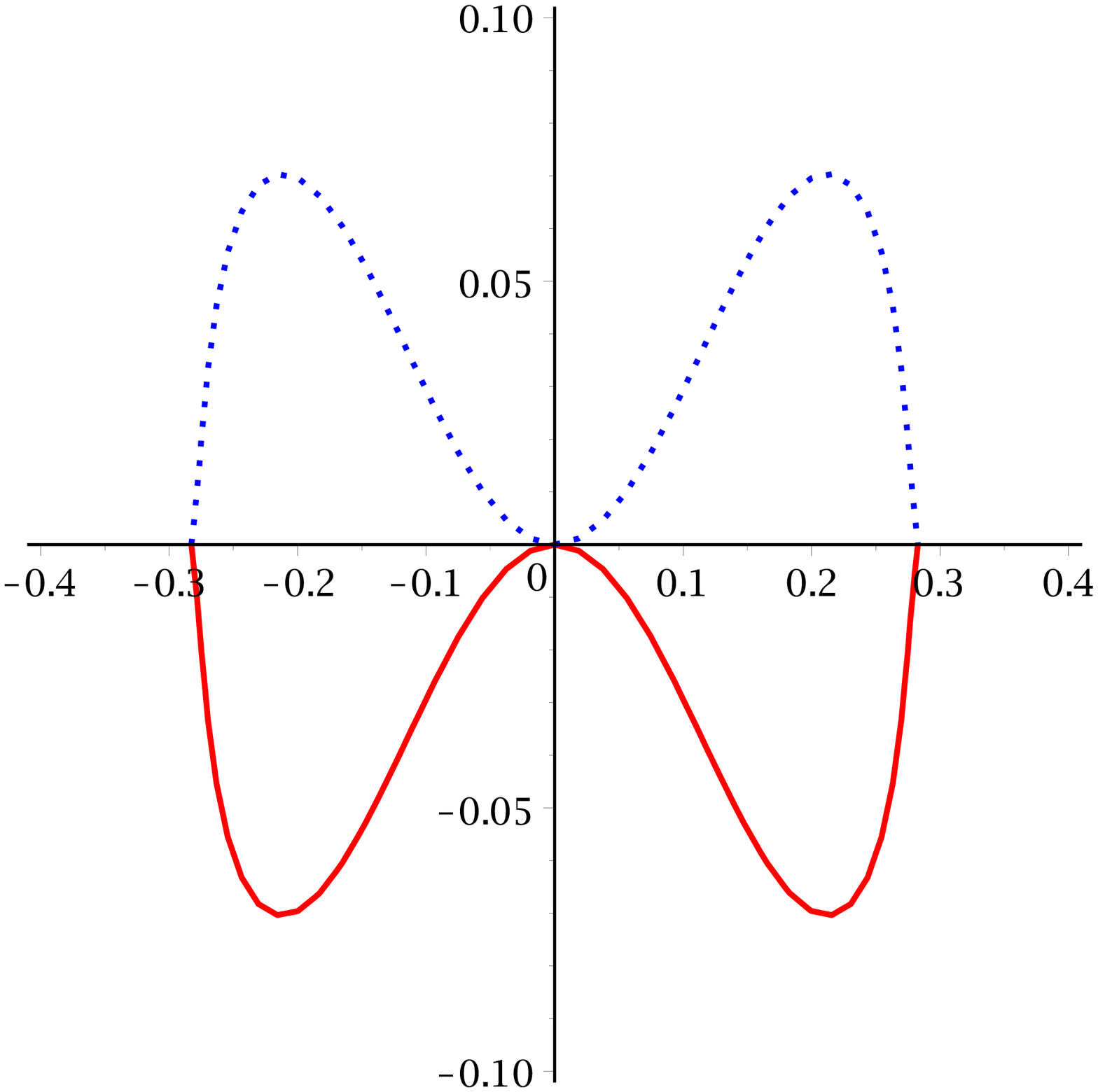}
	}
	\subfigure[$N=0.2$, $a=0.3$ and $v=0.08$]{
		 \includegraphics[width=0.3\textwidth]{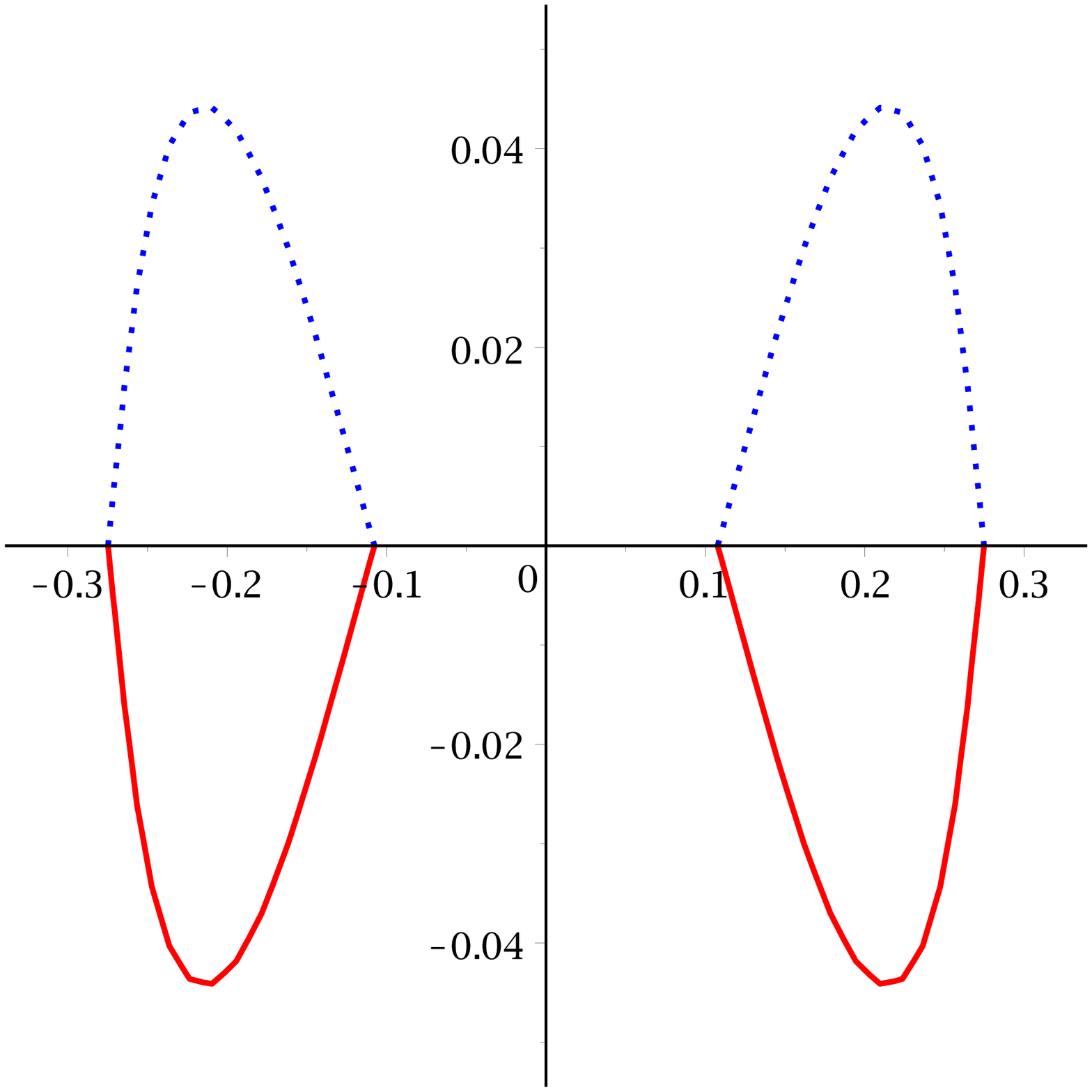}
	}
	\subfigure[$N=0.2$, $a=0.3$ and $v=0.11$]{
		 \includegraphics[width=0.3\textwidth]{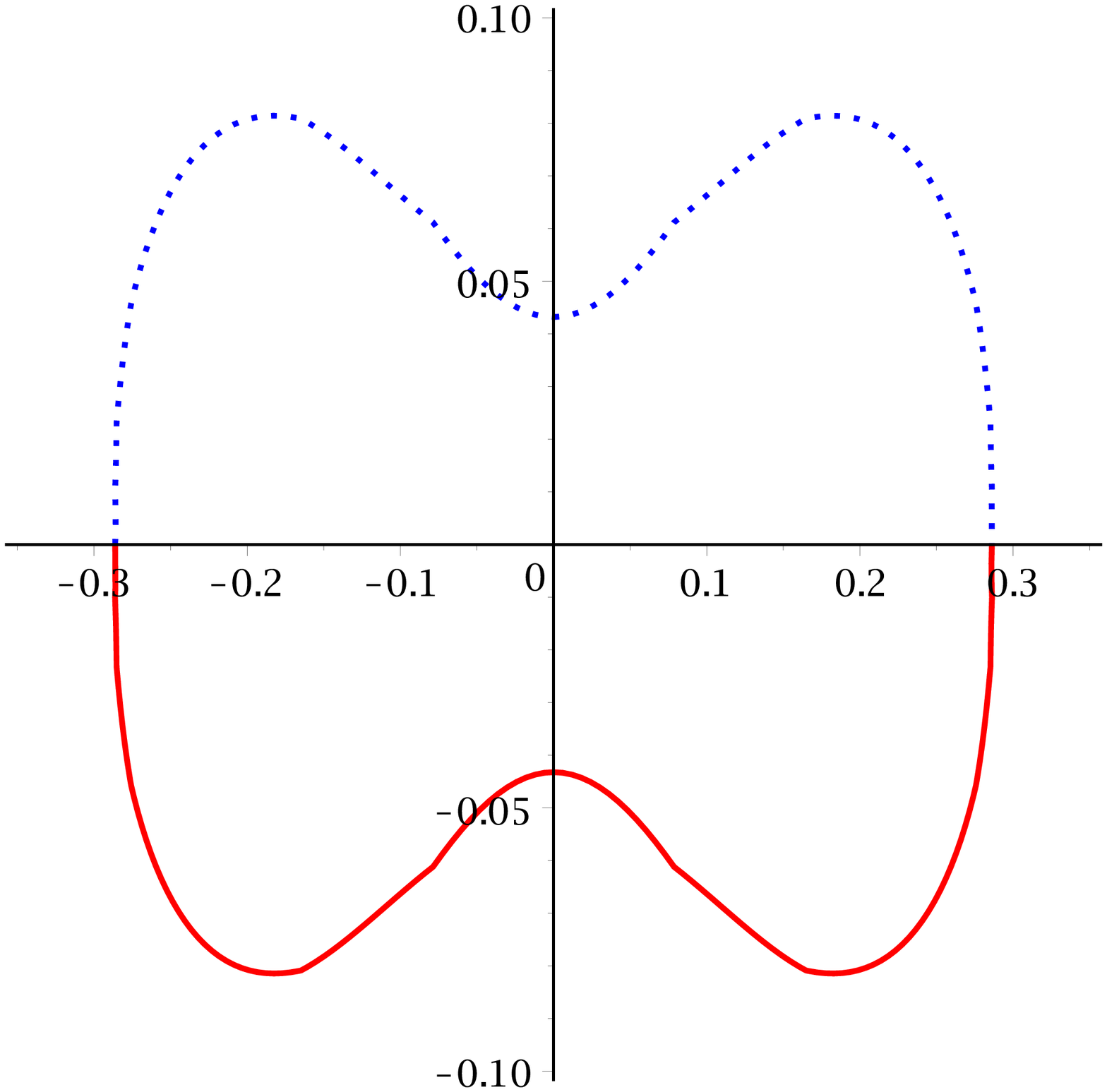}
	}
	\caption{Plots of the singularity given by $B-aA=0$, the dotted blue line represents $r \leq 0$ and the red line $r \geq 0$.}
	\label{pic:singularity-2D}
\end{figure}

Due to the divergence in the Kretschmann scalar, the singularity is defined by $B-aA=r^2+(N_g+a\cos{\vartheta})^2-v^2=0$. The NUT-parameter $N_g$ and the charge-parameter $v$ influence the shape of the singularity in such a way that it varies from ring singularities to three dimensional structures in contrast to the Kerr metric, where only a ring is possible.  We observed this feature in the Einstein-Maxwell-dilaton-axion spacetime \cite{Flathmann:2015} due to the presence of a dilaton charge at negative radial coordinate, but here these structures can also appear for a positive one. This could also lead to a different behaviour of near horizon geodesics affected by the singularity. In Fig. \ref{pic:singularity-2D} a two dimensional projection is shown.

\subsection{The geodesic equations}
We use the Hamilton-Jacobi formalism to derive the geodesic equation of test particles and light. The ansatz for the action
\begin{equation}
	S=\frac{1}{2}\delta \tau -Et+L\varphi + S_r(r) + S_\vartheta(\vartheta) \, ,
\end{equation}
solves the Hamilton-Jacobi equation
\begin{equation}
	\frac{\partial S}{\partial \tau} + \frac{1}{2} g^{\mu\nu} \frac{\partial S}{\partial x^\mu}\frac{\partial S}{\partial x^\nu}=0 \, .
\label{eqn:ham-jac}
\end{equation}
Here $\delta$ is either equal to $0$ for light or equal to $1$ for massive particles, $\tau$ is an affine parameter along the geodesic, which corresponds to the proper time for particles and $E$ and $L$ are the energy and the angular momentum of the test particle.\\
We obtain one differential equation for each coordinate, by separating the Hamilton-Jacobi equation (\ref{eqn:ham-jac}) with the Carter \cite{Carter:1968rr} constant $K$
\begin{eqnarray}
 \left(\frac{\dd \tr}{\dd \gamma}\right)^2 &=& X \, , \label{eqn:r-equation} \\
 \sin^2{\vartheta}\left(\frac{\dd \vartheta}{\dd \gamma}\right)^2 &=& Y \, , \label{eqn:theta-equation} \\
\left(\frac{\dd \phi}{\dd \gamma}\right) &=& \frac{\ta\tXi(\tB E -\ta \tL \tXi)}{\tR} +\frac{\tXi (\tA E -\tL \tXi)}{\tTh \sin^2{\vartheta}} \, , \label{eqn:phi-equation} \\
\left(\frac{\dd \tlt}{\dd \gamma}\right) &=& \frac{\tB(\tB E -\ta\tL\tXi)}{\tR} +\frac{\tA(\tL\tXi-\tA E)}{\tTh \sin^2{\vartheta}} \, . \label{eqn:t-equation}
\end{eqnarray}
Here we used dimensionless quantities 
\begin{equation}
	\tr=\frac{r}{2m} \ , 
	\,\, \tlt=\frac{t}{2m} \ , 
	\,\, \tilde{\tau}=\frac{\tau}{2m} \ , 
	\,\, \tN=\frac{N_g}{2m} \ ,
	\,\, \ta=\frac{a}{2m} \ , \\
	\,\, \tg=2mg \ , 
	\,\, \te=\frac{e}{2m}  \ ,
	\,\, \tv=\frac{b}{2m}  \ ,
	\,\, \tK=\frac{K}{2m}  \ ,
	\,\, \tL=\frac{L}{2m} \ 
\end{equation}
and the Mino-time \cite{Mino:2003yg} $\gamma$ with $\dd \tilde{\tau}=(\tB-\ta \tA) \dd\gamma$. 
We also defined the functions
\begin{equation}
\begin{split}
  X &= (\tB E -\ta\tL\tXi)^2+\tR(\tK -\tB\delta) \, , \\
  Y &= -(\tA E-\tL\tXi)^2+\tTh\sin^2{\vartheta}(\ta\tA\delta-\tK) \, ,\\
  \tR &= \tr^2-\tr +\ta^2+\te^2-\tN^2 \tg^2[\tr^4+(\ta^2+6\tN^2-2\tv^2)\tr^2+3\tN^2(\ta^2-\tN^2)] \, ,\\
  \tTh &= 1 -\ta^2\tg^2\cos^2{\vartheta}+4\ta^2\tg^2\tN\cos{\vartheta} \, , \\
  \tA &= \ta^2 \sin^2{\vartheta}+2 \tN (1-\cos{\vartheta}) \, , \\
  \tB &= \tr^2+(\tN+\ta)^2-\tv^2 \, ,\\
  \tXi &= 1-\ta^2\tg^2-4\ta\tN\tg^2 \, .
\end{split}
\end{equation}

\section{Classification of the geodesic motion}
The characteristics of the geodesic motion are defined by the function $Y$ and the polynomial $X$. We will study their properties with the aim to give a full classification of the possible orbits. In order to obtain real values for $\tr$ and $\vartheta$, eq. (\ref{eqn:r-equation}) and eq. (\ref{eqn:theta-equation}) lead to the conditions $X \geq 0$ and $Y \geq 0$. From the second condition we derive that $\tK \geq \ta\tA\delta$ if $\tTh \geq 0$ and $\tK \leq \ta\tA\delta$ if $\tTh \leq 0$. Due to its shape, the conditions for a particle orbit that ends in the singularity can only be analyzed in special cases.
\subsection{The $\vartheta$-motion}
The zeros of the function $Y$ are the turning points of the latitudinal motion. We substitute $\nu=\cos{\vartheta}$ into $Y$ and analyze the zeros in the interval $\nu=[-1,1]$. The number of zeros changes if $\nu=1$ or $\nu=-1$ is crossed or if $Y$ has a double zero. The first condition $Y(\nu=1)=0$ is only possible if $\tL=0$ and $Y(\nu=-1)=0$ holds if 
\begin{equation}
 \tL=\frac{4E\tN}{\tXi} \, .
\end{equation}
Double zeros appear if
\begin{equation}
	Y(\nu)=0 \quad \text{and} \quad \frac{\dd Y (\nu)}{\dd\nu}=0 \, ,
\end{equation}
We distinguish between four configurations (a)-(d) (compare fig. \ref{pic:parametric-diagrams1} and \ref{pic:parametric-diagrams2}):

\begin{enumerate}
 \item Region (a): Only negative zeros in the interval $\nu \in [-1,1]$. The motion takes place below the equatorial plane.
 \item Region (b): Positive and negative zeros are possible in the interval  $\nu \in [-1,1]$. The motion crosses the equatorial plane.
 \item Region (c): Only positive zeros in the interval $\nu \in [-1,1]$. The motion takes place above the equatorial plane.
 \item Region (d): No zeros in the interval $\nu \in[-1,1]$. Geodesic motion is only possible if $Y>0$ for all $\nu \in[-1,1]$
\end{enumerate}

The results of all conditions will be combined with the $\tr$-motion in parametric $\tL-E^2$-diagrams.

\subsection{The $\tr$-motion}
First of all let us introduce possible orbits for $r_+>r_->0$.
\begin{enumerate}
 \item \textit{Transit orbit} (TrO): $\tr \in (-\infty,\infty)$
 \item \textit{Escape orbit} (EO): $\tr \in (\tr_1 > 0 ,\infty)$ or $\tr \in (-\infty,\tr_1)$ , with $\tr_1 < 0$
 \item \textit{Two-world escape orbit} (TEO): $\tr \in (\tr_1,\infty)$, with $0<\tr_1 < r_-$
 \item \textit{Crossover two-world escape orbit} (CTEO): $\tr \in (\tr_1,\infty)$, with $\tr_1 < 0$
 \item \textit{Bound orbit} (BO): $\tr \in [\tr_1,\tr_2]$, with $\tr_1, \tr_2 > r_+$, $0 < \tr_1 < \tr_2 < r_-$ or $\tr_1,\tr_2 < 0$
 \item \textit{Crossover bound orbit} (CBO): $\tr \in [\tr_1,\tr_2]$, with $\tr_1 < 0$ and $0 < \tr_2 < r_-$
 \item \textit{Many-world bound orbit} (MBO): $\tr \in [\tr_1,\tr_2]$, with $0 < \tr_1 < r_-$ and $r_+ < \tr_2$
 \item \text{Crossover many-world bound orbit} (CMBO): $\tr \in [\tr_1,\tr_2]$, with $\tr_1 < 0$ and $r_+ < \tr_2$
\end{enumerate}

\begin{figure}[ht]
	\centering
	\subfigure[$\vartheta$-motion]{
		\includegraphics[width=0.31\textwidth]{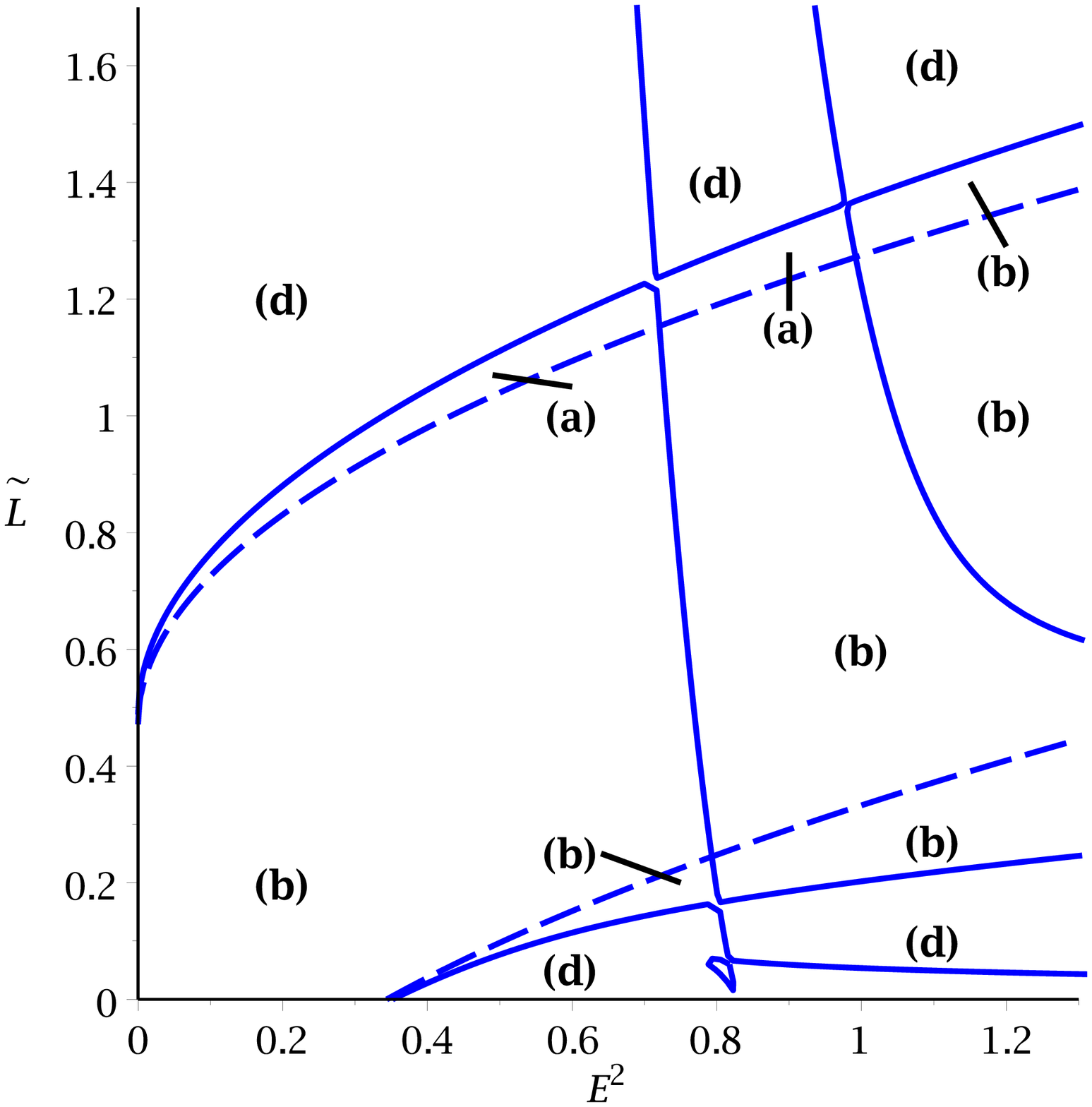}
	}
	\subfigure[$\tr$-motion]{
		\includegraphics[width=0.31\textwidth]{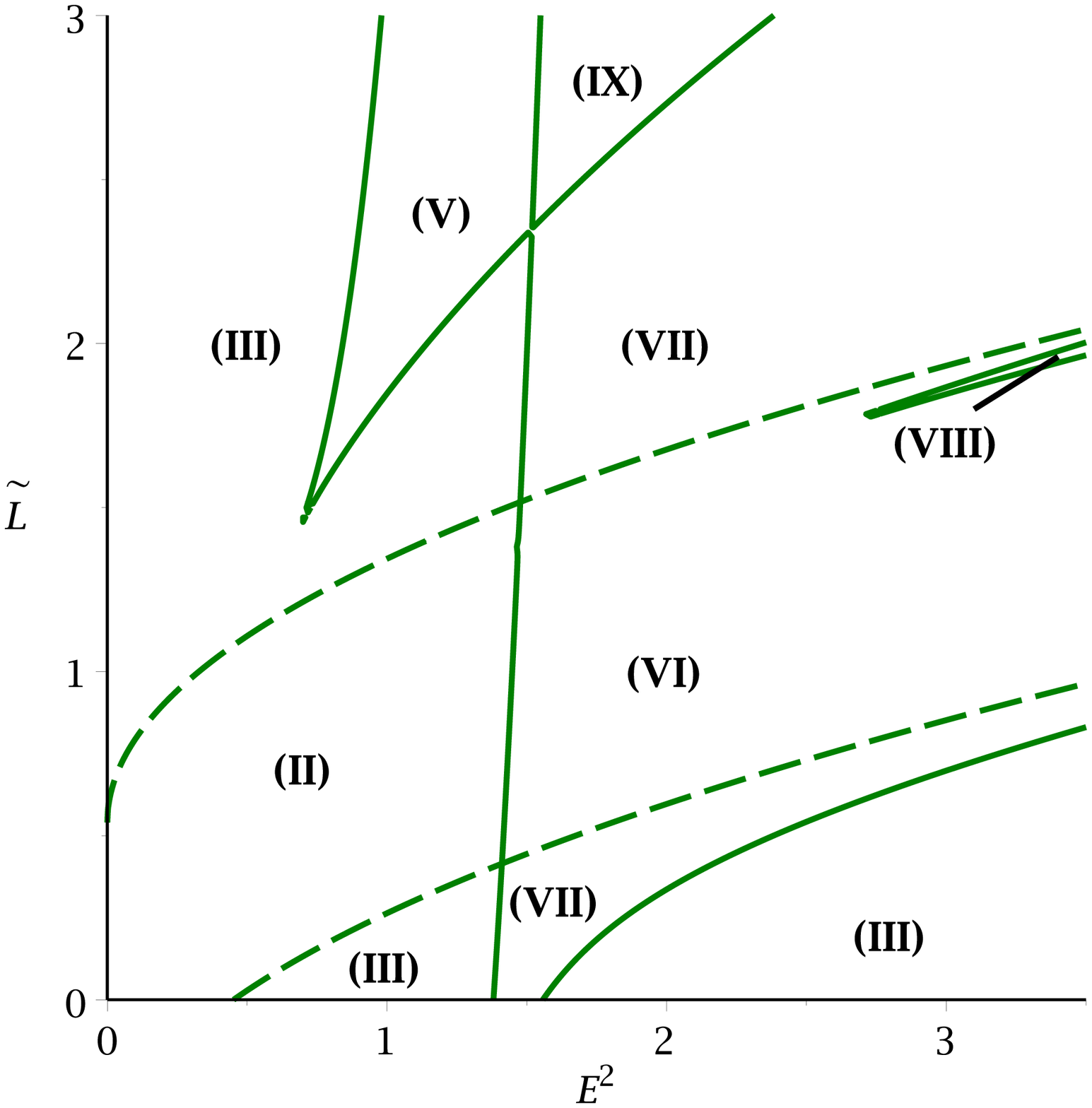}
	}
	\subfigure[Combined diagram]{
		\includegraphics[width=0.31\textwidth]{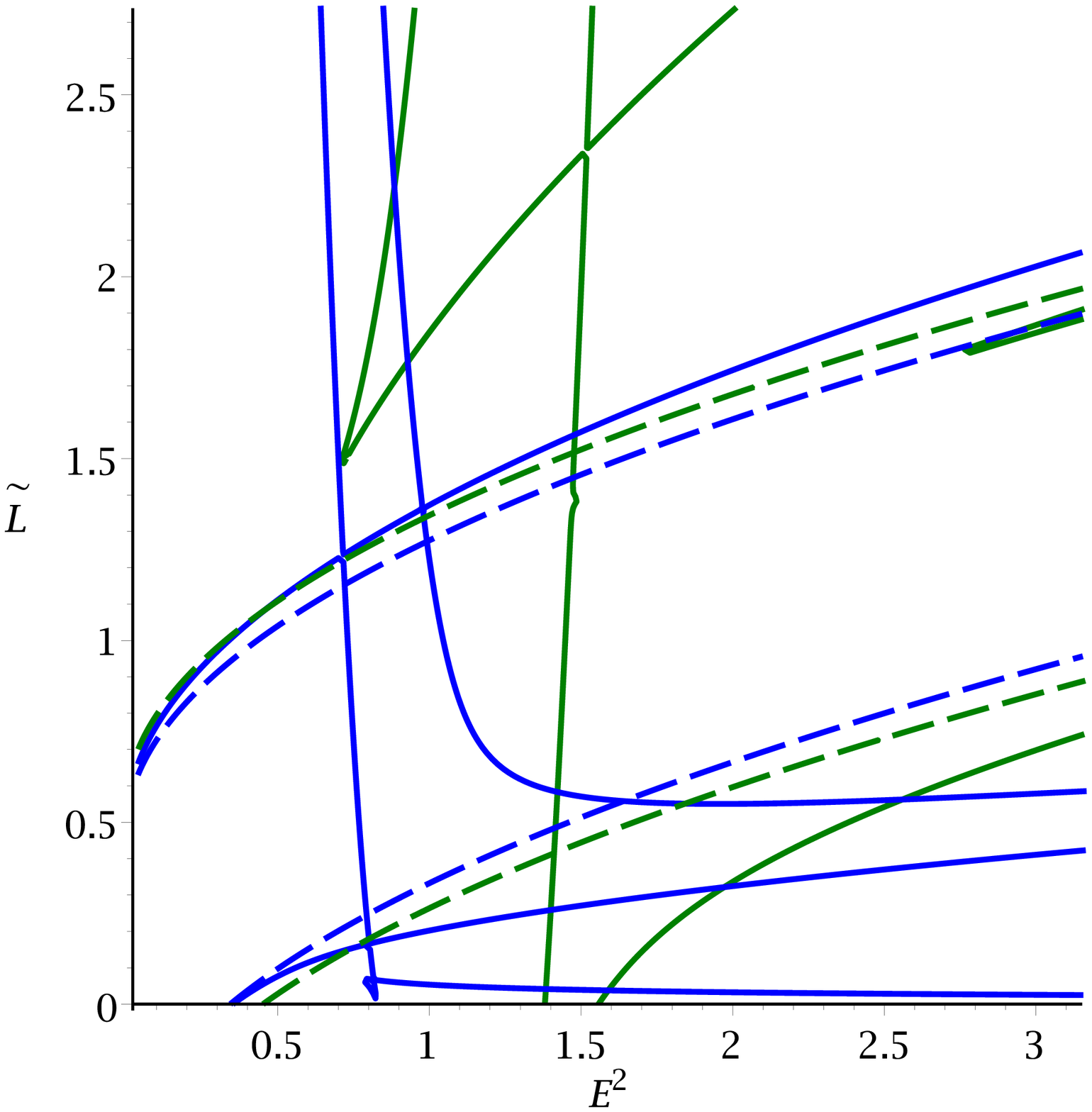}
	}
	\caption{Parametric $\tL-E^2$-diagrams for $\delta=0$, $\ta=0.4$, $\tg=0.1$, $\tN=0.2$, $\te=0.3$, $\tv=0.2$ and $\tK=0.1$. The continuous lines represent double zeros of $Y(\nu)$ in blue and $X(\tr)$ in green. The dashed lines show zeros at $\tr=0$ and $\nu=0$.}
 \label{pic:parametric-diagrams1}
\end{figure}

\begin{figure}[ht]
	\centering
	\subfigure[$\vartheta$-motion]{
		\includegraphics[width=0.31\textwidth]{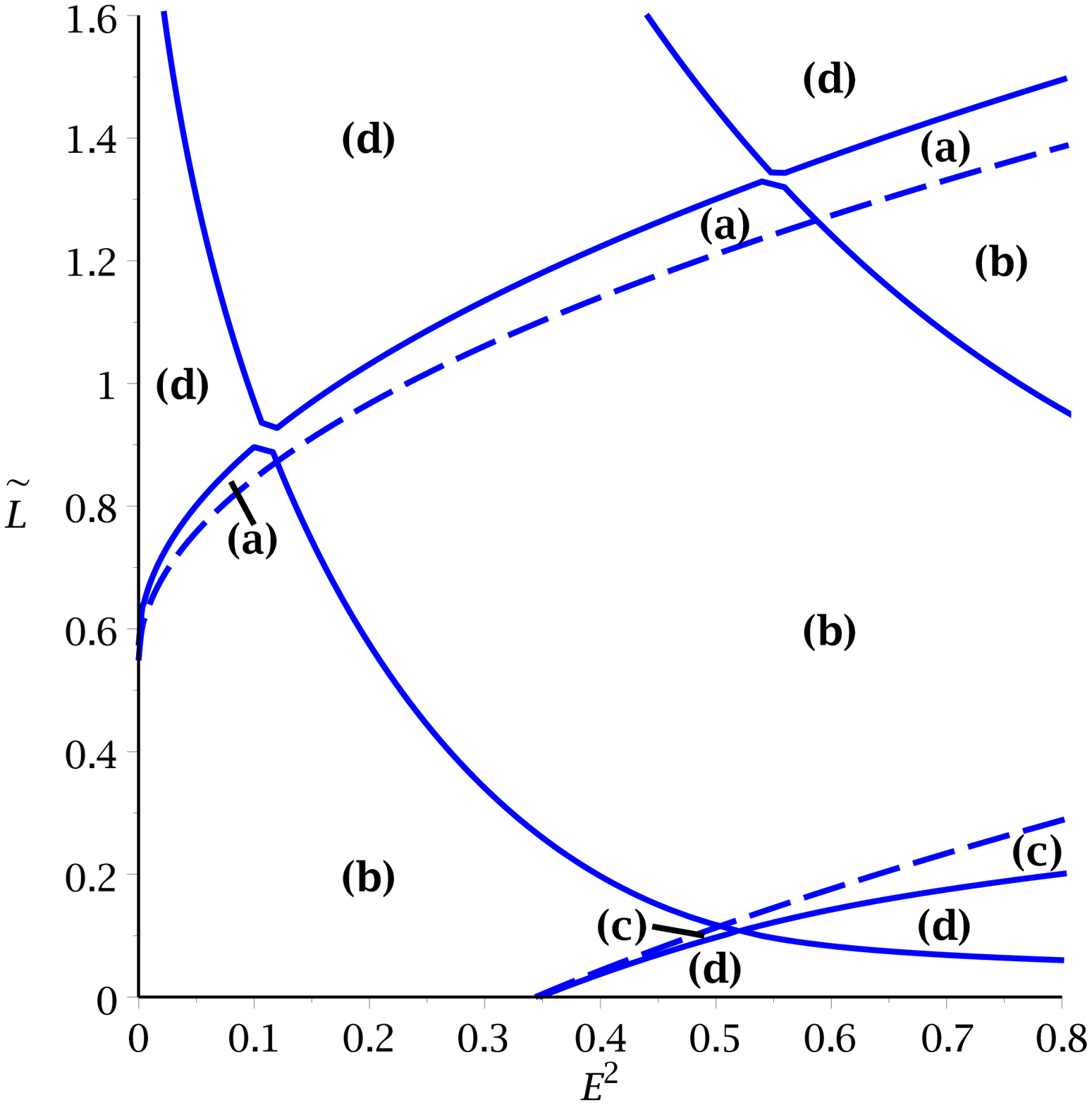}
	}
	\subfigure[$\tr$-motion]{
		\includegraphics[width=0.31\textwidth]{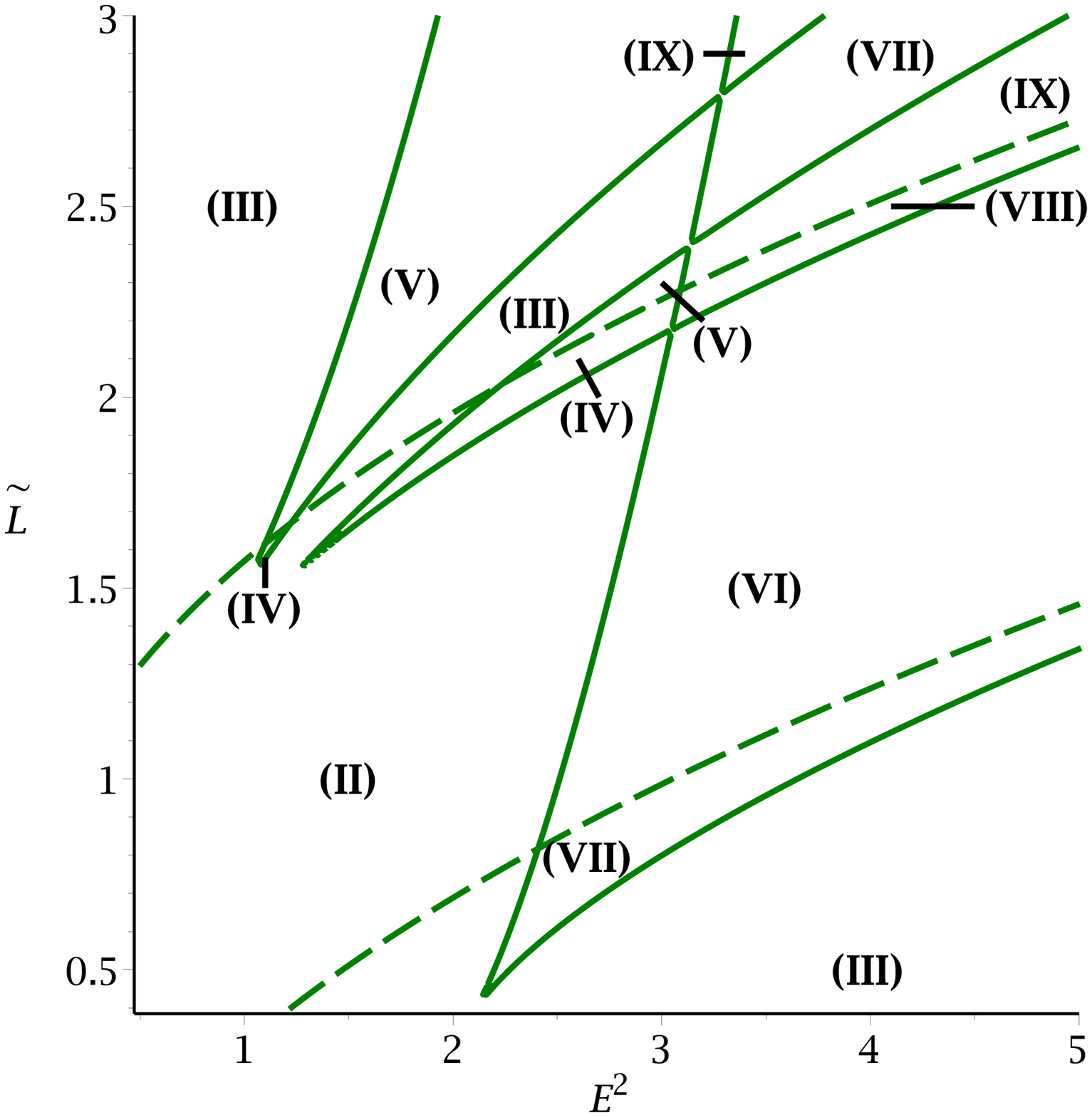}
	}
	\subfigure[Combined diagram]{
		\includegraphics[width=0.31\textwidth]{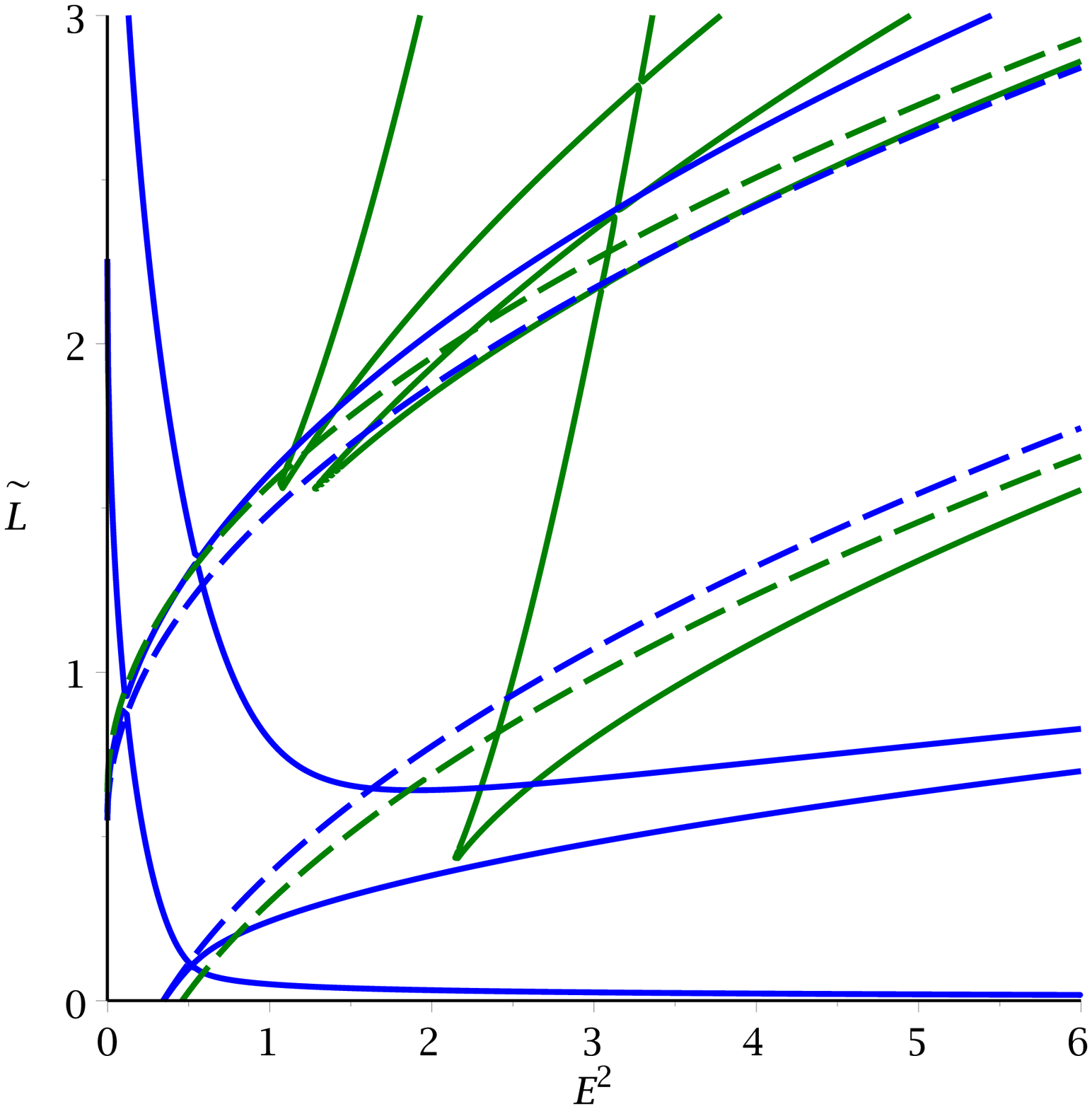}
	}
	\caption{Parametric $\tL-E^2$-diagrams for $\delta=0$, $\ta=0.4$, $\tg=0.55$, $\tN=0.2$, $\te=0.3$, $\tv=0.2$ and $\tK=0.1$. The continuous lines represent double zeros of $Y(\nu)$ in blue and $X(\tr)$ in green. The dashed lines show zeros at $\tr=0$ and $\nu=0$.}
 \label{pic:parametric-diagrams2}
\end{figure}

We can characterize the radial motion by analyzing the zeros of the polynomial $X$, since these are the turning points. As for the $\vartheta$-motion, double zeros appear if
\begin{equation}
	X(\tr)=0 \quad \text{and} \quad \frac{\dd X (\tr)}{\dd\tr}=0 \, ,
\end{equation}
and the number of positive or negative zeros changes, if one zero crosses $\tr=0$.
In figure \ref{pic:parametric-diagrams1} and \ref{pic:parametric-diagrams2} we show two examples of parametric $\tL-E^2$-diagrams for the $\vartheta$- and the $\tr$-motion.


The areas in figure \ref{pic:parametric-diagrams1} and \ref{pic:parametric-diagrams2} are named to distinguish between configurations of the zeros of the polynomial $X$:
\begin{enumerate}
 \item Region (I): $X(\tr)$ has no real zero for $\tr \in (-\infty, \infty)$. Only TrO  are possible.
 \item Region (II): Two real zeros $\tr_1$ and $\tr_2$ in the interval $(-\infty, \infty)$. Here MBO are possible.
 \item Region (III): Two real zeros $\tr_1$ and $\tr_2$ with $\tr_1 \in (-\infty, 0]$ and $\tr_1 \in [0, \infty)$. The new class of a CMBO is possible.
 \item Region (IV): Four positive real zeros exists. In this region BO and MBO exist.
 \item Region (V): Four real zeros $\tr_1$, $\tr_2$, $\tr_3$, $\tr_4$ with $\tr_1 \in (-\infty,0] $ and $\tr_2, \tr_3, \tr_4 \in [0, \infty)$. Here CBO and MBO are possible.
 \item Region (VI): Two positive and two negative real zeros. Here BO and MBO exist.
 \item Region (VII): One positive and three negative real zeros. BO and MBO are possible.
 \item Region (VIII): Six real zeros $\tr_1,\tr_2, \tr_3, \tr_4, \tr_5, \tr_6$ with $\tr_1, \tr_2, \tr_3, \tr_4 \in (-\infty,0]$ and $\tr_5, \tr_6 \in [0,\infty)$. Here BO and MBO are possible.
 \item Region (IX): Three positive and three negative real zeros. Possible orbits are a BO and a CBO which could lie inside the singularity.    
\end{enumerate}

\begin{figure}[!ht]
	\centering
	\subfigure[$\delta=0$, $\ta=0.4$, $\tg=0$, $\tN=0$, $\te=0$, $\tv=0$, $\tK=-1$ and $\tL=4.6$]{
		\includegraphics[width=0.31\textwidth]{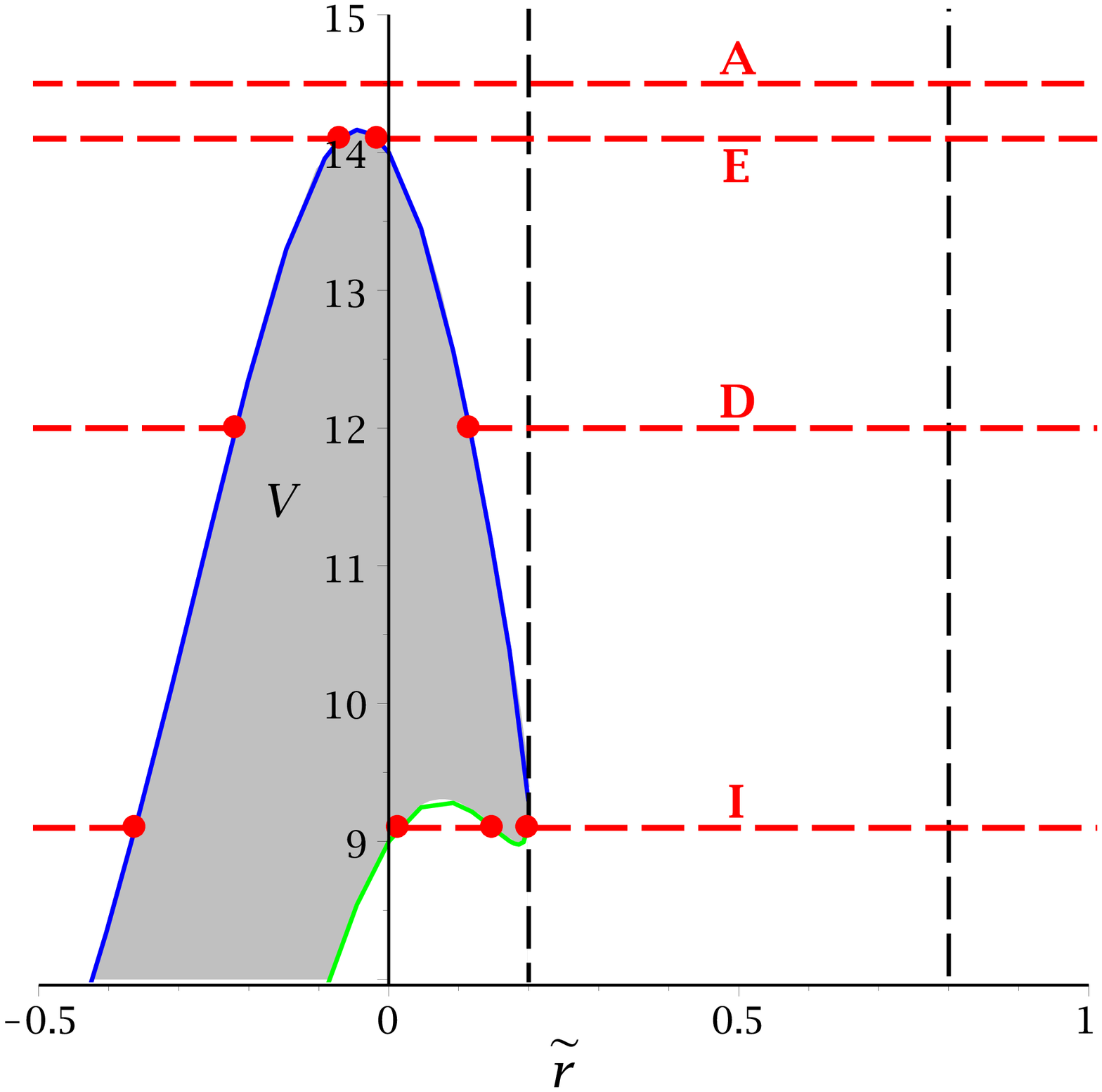}
	}
	\subfigure[$\delta=1$, $\ta=0.4$, $\tg=0$, $\tN=0$, $\te=0$, $\tv=0$, $\tK=-10$ and $\tL=2.3$]{
		\includegraphics[width=0.31\textwidth]{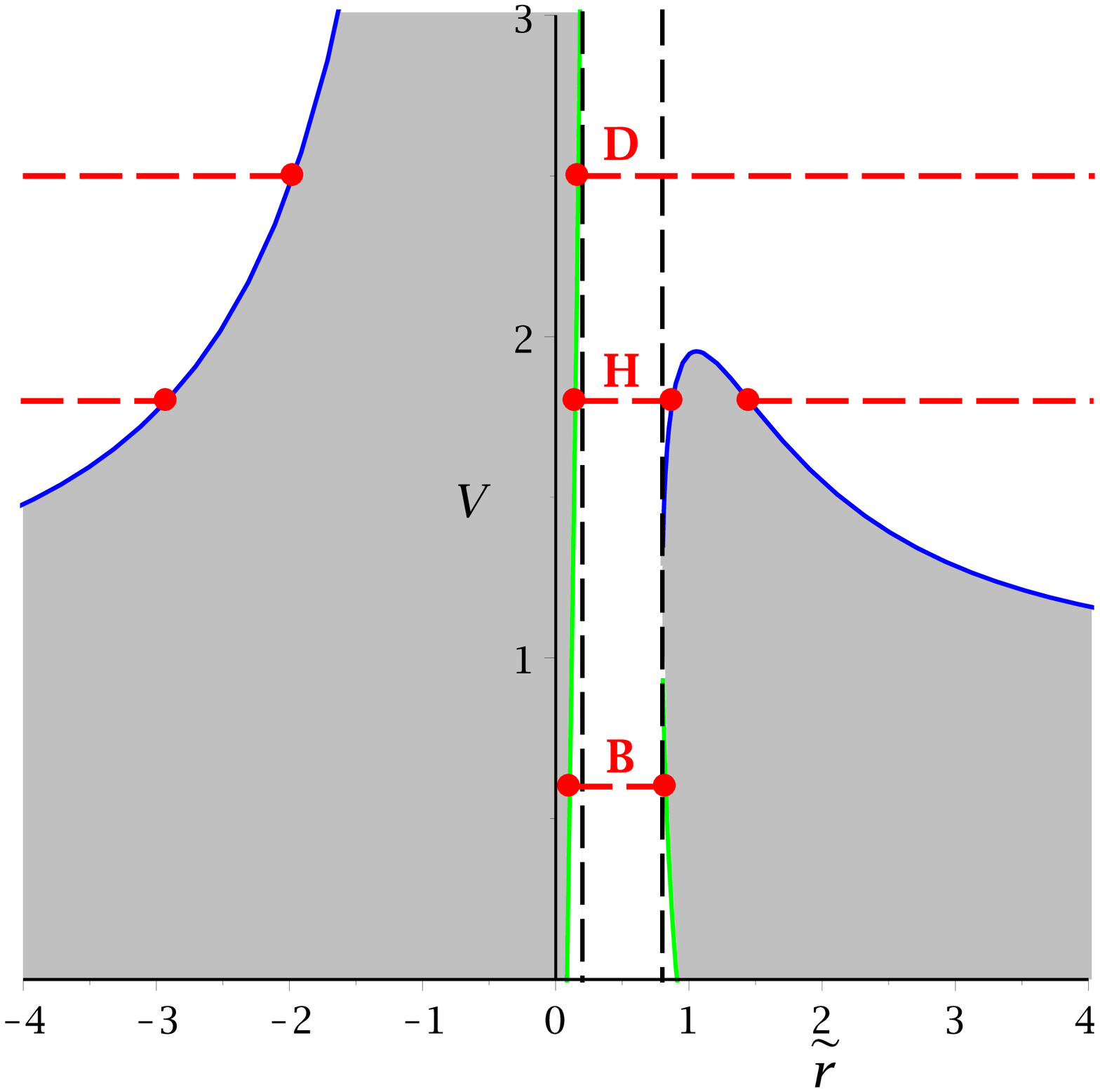}
	}
	\subfigure[$\delta=0$, $\ta=0.4$, $\tg=2$, $\tN=\sqrt{0.1}$, $\te=0.38$, $\tv=1.3$, $\tK=-1$ and $\tL=1.3$]{
		\includegraphics[width=0.31\textwidth]{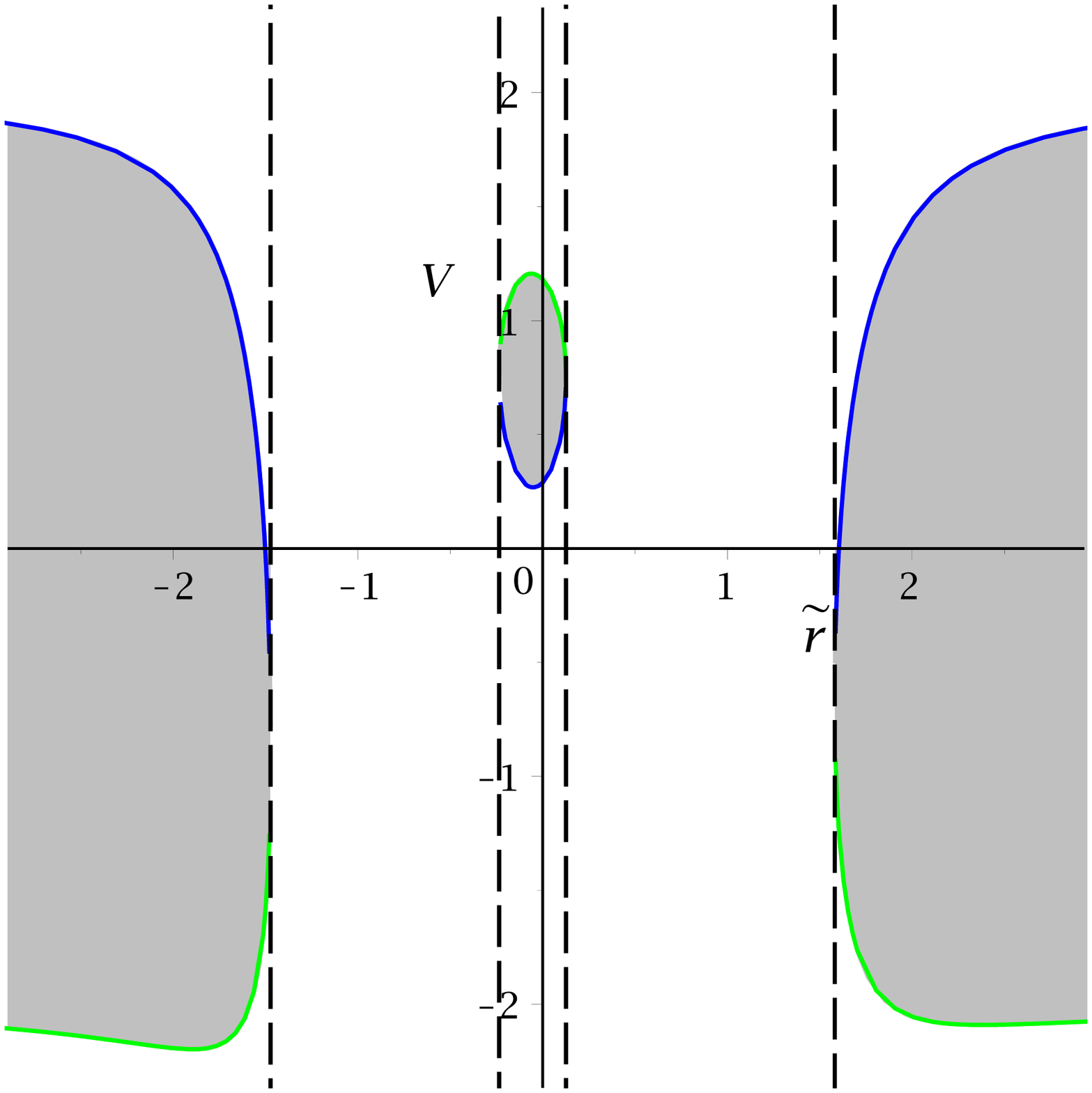}
	}
	\caption{Illustration of the effective potential $V_{\pm}$ and energies for examples of orbit types of table \ref{tab:orbit-types}. The blue line represents $V_+$ and the green line $V_-$. The red dots denote the turning points of the particle and the grey area forbids geodesic motion due to $X\le 0$.}
 \label{pic:potential1}
\end{figure}

In addition we can rewrite $X$ as

\begin{equation}
 X = f(\tr)(E-V_+)(E-V_-)
\end{equation}

to determine an effective potential $V_\pm$. Now we can visualize the turning points of the $\tr$-motion with the intersections of $E$ and $V_\pm$. Figure \ref{pic:potential1}(a) and \ref{pic:potential1}(b) show two examples of the effective potential $V_{\pm}$, where orbits are shown with vanishing cosmological constant. In figure \ref{pic:potential1}(c) we can see, that negative horizons have the same effect on the effective potential, as the horizons at positive $\tr$. We assume that similar orbits are possible and only take a look at the particle motion for two positive horizons.  
To summarize, we present a list of all possible orbit types in table \ref{tab:orbit-types}.

\begin{table}[!ht]
\begin{center}
\begin{tabular}{|lccll|}\hline
type &  zeros & region  & range of $\tr$ & orbit \\
\hline\hline
A &  0 &  &
\begin{pspicture}(-4,-0.2)(3.5,0.2)
\psline[linewidth=0.5pt]{->}(-4,0)(3.5,0)
\psline[linewidth=0.5pt](-2,-0.2)(-2,0.2)
\psline[linewidth=0.5pt,doubleline=true](0.3,-0.2)(0.3,0.2)
\psline[linewidth=0.5pt,doubleline=true](1.2,-0.2)(1.2,0.2)
\psline[linewidth=1.2pt]{-}(-4,0)(3.5,0)
\end{pspicture}
  & TrO
\\  \hline
B & 2 & II a,b,c  &
\begin{pspicture}(-4,-0.2)(3.5,0.2)
\psline[linewidth=0.5pt]{->}(-4,0)(3.5,0)
\psline[linewidth=0.5pt](-2,-0.2)(-2,0.2)
\psline[linewidth=0.5pt,doubleline=true](0.3,-0.2)(0.3,0.2)
\psline[linewidth=0.5pt,doubleline=true](1.2,-0.2)(1.2,0.2)
\psline[linewidth=1.2pt]{*-*}(0,0)(1.5,0)
\end{pspicture}
& MBO
\\  \hline
C & 2 & III a,c  &
\begin{pspicture}(-4,-0.2)(3.5,0.2)
\psline[linewidth=0.5pt]{->}(-4,0)(3.5,0)
\psline[linewidth=0.5pt](-2,-0.2)(-2,0.2)
\psline[linewidth=0.5pt,doubleline=true](0.3,-0.2)(0.3,0.2)
\psline[linewidth=0.5pt,doubleline=true](1.2,-0.2)(1.2,0.2)
\psline[linewidth=1.2pt]{*-*}(-2.5,0)(1.5,0)
\end{pspicture}
& CMBO 
\\  \hline
D   & 2 &  &
\begin{pspicture}(-4,-0.2)(3.5,0.2)
\psline[linewidth=0.5pt]{->}(-4,0)(3.5,0)
\psline[linewidth=0.5pt](-2,-0.2)(-2,0.2)
\psline[linewidth=0.5pt,doubleline=true](0.3,-0.2)(0.3,0.2)
\psline[linewidth=0.5pt,doubleline=true](1.2,-0.2)(1.2,0.2)
\psline[linewidth=1.2pt]{-*}(-4,0)(-2.5,0)
\psline[linewidth=1.2pt]{*-}(0,0)(3.5,0)
\end{pspicture}
& EO, TEO 
\\ \hline
E  & 2 &  &
\begin{pspicture}(-4,-0.2)(3.5,0.2)
\psline[linewidth=0.5pt]{->}(-4,0)(3.5,0)
\psline[linewidth=0.5pt](-2,-0.2)(-2,0.2)
\psline[linewidth=0.5pt,doubleline=true](0.3,-0.2)(0.3,0.2)
\psline[linewidth=0.5pt,doubleline=true](1.2,-0.2)(1.2,0.2)
\psline[linewidth=1.2pt]{-*}(-4,0)(-3.5,0)
\psline[linewidth=1.2pt]{*-}(-2.5,0)(3.5,0)
\end{pspicture}
  & EO, CTEO
\\ \hline
F &  4 & V a &
\begin{pspicture}(-4,-0.2)(3.5,0.2)
\psline[linewidth=0.5pt]{->}(-4,0)(3.5,0)
\psline[linewidth=0.5pt](-2,-0.2)(-2,0.2)
\psline[linewidth=0.5pt,doubleline=true](0.3,-0.2)(0.3,0.2)
\psline[linewidth=0.5pt,doubleline=true](1.2,-0.2)(1.2,0.2)
\psline[linewidth=1.2pt]{*-*}(0,0)(1.5,0)
\psline[linewidth=1.2pt]{*-*}(-2.5,0)(-0.5,0)
\end{pspicture}
& CBO, MBO 
\\
F$_+$ &  & IV a,b  &
\begin{pspicture}(-4,-0.2)(3.5,0.2)
\psline[linewidth=0.5pt]{->}(-4,0)(3.5,0)
\psline[linewidth=0.5pt](-2,-0.2)(-2,0.2)
\psline[linewidth=0.5pt,doubleline=true](0.3,-0.2)(0.3,0.2)
\psline[linewidth=0.5pt,doubleline=true](1.2,-0.2)(1.2,0.2)
\psline[linewidth=1.2pt]{*-*}(0,0)(1.5,0)
\psline[linewidth=1.2pt]{*-*}(-1.5,0)(-0.5,0)
\end{pspicture}
& BO, MBO 
\\
F$_-$ &  & VI a,b,c  &
\begin{pspicture}(-4,-0.2)(3.5,0.2)
\psline[linewidth=0.5pt]{->}(-4,0)(3.5,0)
\psline[linewidth=0.5pt](-2,-0.2)(-2,0.2)
\psline[linewidth=0.5pt,doubleline=true](0.3,-0.2)(0.3,0.2)
\psline[linewidth=0.5pt,doubleline=true](1.2,-0.2)(1.2,0.2)
\psline[linewidth=1.2pt]{*-*}(0,0)(1.5,0)
\psline[linewidth=1.2pt]{*-*}(-3.5,0)(-2.5,0)
\end{pspicture}
& BO, MBO 
\\ \hline
G & 4 & IV a &
\begin{pspicture}(-4,-0.2)(3.5,0.2)
\psline[linewidth=0.5pt]{->}(-4,0)(3.5,0)
\psline[linewidth=0.5pt](-2,-0.2)(-2,0.2)
\psline[linewidth=0.5pt,doubleline=true](0.3,-0.2)(0.3,0.2)
\psline[linewidth=0.5pt,doubleline=true](1.2,-0.2)(1.2,0.2)
\psline[linewidth=1.2pt]{*-*}(0,0)(1.5,0)
\psline[linewidth=1.2pt]{*-*}(2,0)(3,0)
\end{pspicture}
  &  BO, MBO
\\ \hline
H  & 4 &  &
\begin{pspicture}(-4,-0.2)(3.5,0.2)
\psline[linewidth=0.5pt]{->}(-4,0)(3.5,0)
\psline[linewidth=0.5pt](-2,-0.2)(-2,0.2)
\psline[linewidth=0.5pt,doubleline=true](0.3,-0.2)(0.3,0.2)
\psline[linewidth=0.5pt,doubleline=true](1.2,-0.2)(1.2,0.2)
\psline[linewidth=1.2pt]{-*}(-4,0)(-2.5,0)
\psline[linewidth=1.2pt]{*-*}(0,0)(1.5,0)
\psline[linewidth=1.2pt]{*-}(2,0)(3.5,0)
\end{pspicture}
  & EO, MBO, EO
\\ \hline
I  & 4 &  &
\begin{pspicture}(-4,-0.2)(3.5,0.2)
\psline[linewidth=0.5pt]{->}(-4,0)(3.5,0)
\psline[linewidth=0.5pt](-2,-0.2)(-2,0.2)
\psline[linewidth=0.5pt,doubleline=true](0.3,-0.2)(0.3,0.2)
\psline[linewidth=0.5pt,doubleline=true](1.2,-0.2)(1.2,0.2)
\psline[linewidth=1.2pt]{-*}(-4,0)(-2.5,0)
\psline[linewidth=1.2pt]{*-*}(-1.5,0)(-0.5,0)
\psline[linewidth=1.2pt]{*-}(0,0)(3.5,0)
\end{pspicture}
  & EO, BO, TEO
\\ \hline
J & 4 & VII a,b,c  &
\begin{pspicture}(-4,-0.2)(3.5,0.2)
\psline[linewidth=0.5pt]{->}(-4,0)(3.5,0)
\psline[linewidth=0.5pt](-2,-0.2)(-2,0.2)
\psline[linewidth=0.5pt,doubleline=true](0.3,-0.2)(0.3,0.2)
\psline[linewidth=0.5pt,doubleline=true](1.2,-0.2)(1.2,0.2)
\psline[linewidth=1.2pt]{*-*}(-2.5,0)(1.5,0)
\psline[linewidth=1.2pt]{*-*}(-3.5,0)(-3,0)
\end{pspicture}
& BO, CMBO
\\ \hline
K & 6 & VIII a,b  &
\begin{pspicture}(-4,-0.2)(3.5,0.2)
\psline[linewidth=0.5pt]{->}(-4,0)(3.5,0)
\psline[linewidth=0.5pt](-2,-0.2)(-2,0.2)
\psline[linewidth=0.5pt,doubleline=true](0.3,-0.2)(0.3,0.2)
\psline[linewidth=0.5pt,doubleline=true](1.2,-0.2)(1.2,0.2)
\psline[linewidth=1.2pt]{*-*}(0,0)(1.5,0)
\psline[linewidth=1.2pt]{*-*}(-1.5,0)(-0.5,0)
\psline[linewidth=1.2pt]{*-*}(-3.5,0)(-2.5,0)
\end{pspicture}
& BO, BO, MBO 
\\ \hline
L & 6 & IX a  &
\begin{pspicture}(-4,-0.2)(3.5,0.2)
\psline[linewidth=0.5pt]{->}(-4,0)(3.5,0)
\psline[linewidth=0.5pt](-2,-0.2)(-2,0.2)
\psline[linewidth=0.5pt,doubleline=true](0.3,-0.2)(0.3,0.2)
\psline[linewidth=0.5pt,doubleline=true](1.2,-0.2)(1.2,0.2)
\psline[linewidth=1.2pt]{*-*}(0,0)(-0.5,0)
\psline[linewidth=1.2pt]{*-*}(-3.5,0)(-3,0)
\psline[linewidth=1.2pt]{*-*}(-2.5,0)(-1.5,0)
\end{pspicture}
& BO, CBO, BO 
\\ \hline\hline
\end{tabular}
\caption{Possible types of orbits in the $\text{U(1)}^2$ dyonic rotating black hole spacetime. The thick lines represent the range of $\tr$. $\tr=0$ is represented by one and the horizons by two vertical lines. The turning points are indicated by thick dots.}
\label{tab:orbit-types}
\end{center}
\end{table}

\section{Solution of the geodesic equations}
In this section we present the analytical solution of the equations (\ref{eqn:r-equation})-(\ref{eqn:t-equation}) in two cases.
The first ones are valid for vanishing gauge coupling constant $\tg$ or for light ($\delta=0$) and are given in terms of the Weierstra{\ss} $\wp$, $\sigma$ and $\zeta$ functions. The second ones can be used for any values of the parameters, but are more complicated to handle. These solutions are given in terms of the Kleinian $\sigma$ function. 

\subsection{The $\tr$-equation}
In general the right hand side of (\ref{eqn:r-equation}) is an order six polynomial $X = \sum_{i=1}^6 a_i\tr^i$ with the coefficients
\begin{equation}
 \begin{split}
a_6 &= -\tg^2\delta  \\
a_5 &= 0 \\
a_4 &= E^2-\delta +\tg^2[\tK+\delta(-7\tN^2-2\tN\ta-2\ta^2+3\tv^2)] \\
a_3 &= \delta \\
a_2 &= 2E\tL\ta(4\tN\ta\tg^2-1)+2E^2(\ta+\tN+\tv)(\ta+\tN-\tv) \\
    &-\delta[((\tN+\ta)^2-\tv^2)(1+\tg^2(6\tN^2+\ta^2-2\tv^2))-\tN^2+3\tg^2\tN^2(\ta^2-\tN^2)+\ta^2+\te^2]\\
    &+\tK(1+\tg^2(6\tN^2+\ta^2-2\tv^2)) \\
a_1 &= \delta(\ta+\tN+\tv)(\ta+\tN-\tv)-\tK \\
a_0 &= \tL^2\ta^2(4\tN\ta\tg^2+\ta\tg^2-1)^2+8E\tL\ta(\tN\ta\tg^2+\frac{1}{4}\ta^2\tg^2-\frac{1}{4})(\ta+\tN+\tv)(\ta+\tN-\tv)\\
    &+((\tN+\ta)^2-\tv^2)^2-\delta(\tN+\ta)^2-\tv^2)(-\tN^2+3\tg^2\tN^2(\ta^2-\tN^2)+\ta^2+\te^2)\\
    &+\tK(-\tN^2+3\tg^2\tN^2(\ta^2-\tN^2)+\ta^2+\te^2)
 \end{split}
\end{equation}
\subsubsection{Elliptic case}
For $\tg$ or $\delta$ equal to zero $a_6=0$ and therefore $X$ is a polynomial of order 4. With the substitution $\tr=\pm \frac{1}{x}+r_X$, where $r_X$ is a zero of $X$, equation (\ref{eqn:r-equation}) becomes
\begin{equation}
 \left(\frac{\dd x}{\dd \gamma}\right)^2=\sum_{i=0}^{3}b_i x^i 
\end{equation}
with fixed coefficients. With $x=\frac{1}{b_3}\left( 4y-\frac{b_2}{3}\right)$ eq. (\ref{eqn:r-equation}) becomes
\begin{equation}
	\left(\frac{\dd y}{\dd\gamma}\right)^2=4y^3-g_2^{\tr}y-g_3^{\tr}= P_3^{\tr} (y) \, ,
	\label{eqn:weierstrass-form}
\end{equation} 
with
\begin{equation}
	g_2^{\tr}=\frac{b_2^2}{12} - \frac{b_1b_3}{4} \, , \qquad  g_3^{\tr}=\frac{b_1b_2b_3}{48} - \frac{b_0b_3^2}{16}-\frac{b_2^3}{216} \ .
\end{equation}

Equation (\ref{eqn:weierstrass-form}) is an elliptic differential equation in the standard Weierstra{\ss} form which can be solved by \cite{Markushevich:1967}
\begin{equation}
	y(\gamma) = \wp\left(\gamma - \gamma'_{\rm in}; g_2^{\tr}, g_3^{\tr}\right) \ ,
\end{equation}
where $\gamma'_{\rm in}$ only depends on the initial value $\tr_{\rm in}$ by $\gamma'_{\rm in}=\gamma_{\rm in}+\int^\infty_{y_{\rm in}}{\frac{\dd y}{\sqrt{4y^3-g_2^{\tr}y-g_3^{\tr}}}}$ with $y_{\rm in}=\pm\frac{b_3}{4\tr_{\rm in}} + \frac{b_2}{12}$. With resubstitution we can obtain the solution of eq. (\ref{eqn:r-equation})
\begin{equation}
	\tr=\pm \frac{b_3}{4 \wp\left(\gamma - \gamma'_{\rm in}; g_2^{\tr}, g_3^{\tr}\right) - \frac{b_2}{3}} +\tr_R\ .
\end{equation}   
\subsubsection{Hyperelliptic case}
Now we look at eq. \ref{eqn:r-equation} for non-vanishing $\tg$ and massive particles. We can transform this equation to
\begin{equation}
 \left(x\frac{\dd x}{\dd \gamma}\right)^2=\sum_{i=0}^5 b_ix^i=:P_5^{\tr}(x)
\end{equation}
with $\tr=\pm \frac{1}{x}+r_X$. Separation leads to  
\begin{equation}
 \gamma-\gamma_{\rm in}=\int_{x_{\rm in}}^x{\frac{x'dx'}{\sqrt{P_5^{\tr}(x')}}}
\label{eqn:int-r-equation}
\end{equation}
Eq. (\ref{eqn:int-r-equation}) is a hyperelliptic integral of the first kind and can be solved in terms of derivatives of the Kleinian $\sigma$-function. These methods were developed in \cite{Hackmann:2008}.
\begin{equation}
 x= -\frac{\sigma_1(\vec{\gamma}_{\infty})}{\sigma_2(\vec{\gamma}_{\infty})} \, , 
\end{equation}
where $\sigma_i=\frac{\partial \sigma(\vec{z})}{\partial z_i}$ and
\begin{equation}
 \vec{\gamma}_{\infty} = \left(-\int_{x}^{\infty}{\frac{\dd x}{\sqrt{P_5^{\tr}(x)}}}, \gamma - \gamma_{\rm in}- \int_{x_{\rm in}}^{\infty}{{\frac{x \dd x}{\sqrt{P_5^{\tr}(x)}}}}\right)^T
\end{equation}
By resubstitution we get the full solution of (\ref{eqn:r-equation})
\begin{equation}
 \tr(\gamma)= \mp \frac{\sigma_2(\vec{\gamma}_{\infty})}{\sigma_1(\vec{\gamma}_{\infty})}+\tilde{r}_X
\end{equation}
\subsection{$\vartheta$-equation}
Eq. (\ref{eqn:theta-equation}) can be solved similar to the $\tr$-equation (\ref{eqn:r-equation}).
By introducing $\nu=\cos{\vartheta}$ eq. (\ref{eqn:theta-equation}) becomes
\begin{equation}
 \left(\frac{\dd \nu}{\dd \gamma}\right)^2 = \sum_{i=0}^6a'_i\nu^i = Y_{\nu}(\nu)
\end{equation}
with the coefficients
\begin{equation}
 \begin{split}
a'_6 &= -\delta\tg^2\ta^4  \\
a'_5 &= -6\delta\tN\tg^2\ta^4 \\
a'_4 &= \delta\tg^2\ta^4-E^2\ta^2-\tK\ta^2\tg^2-\delta\ta(8\ta\tN^2\tg^2-\ta^2\tg^2(\ta+2\tN)-\ta) \\
a'_3 &= 6\delta\tg^2\ta^3\tN-4E^2\tN\ta-4\tK\tN\tg^2-\ta\delta(-4\ta\tN\tg^2(\ta+2\tN)-2\tN) \\
a'_2 &= \tK(1+\ta^2\tg^2)+\ta\delta(8\tN^2\ta\tg^2-\ta-(\ta+2\tN)(1+\ta^2\tg^2)) \\
a'_1 &= 4\tK\ta\tN\tg^2-\ta\delta(4\ta\tN\tg^2(\ta+2\tN)-2\tN)+4\tN E^2(\ta+2\tN)-4\tN\tL E(1-4\tN\ta\tg^2-\ta^2\tg^2) \\
a'_0 &= \ta\delta(\ta+2\tN)-\tK+(E(\ta+2\tN)-L(1-4\tN\ta\tg^2+\ta^2\tg^2))^2
 \end{split}
\end{equation}
\subsubsection{Elliptic case}
As for the $\tr$-equation (\ref{eqn:r-equation}) we can first set $\tg$ or $\delta$ to zero to get an elliptic differential equation of the first kind. The substitution $\nu=\pm\frac{1}{\mu}+\nu_{Y}$, where $\nu_{Y}$ is a zero of $Y_{\nu}$ leads to the form
\begin{equation}
 \left(\frac{\dd \mu}{\dd \gamma}\right)^2 = \sum_{i=0}^3 b'_i \mu^i
\end{equation}
Finally $\mu =\frac{1}{b'_3}\left( 4\xi-\frac{b'_2}{3}\right)$ transforms (\ref{eqn:theta-equation}) into the standard Weierstra{\ss} form
\begin{equation}
	\left(\frac{\dd \xi}{\dd \gamma}\right)^2 = 4\xi^3-g_2^{\vartheta} \xi-g_3^{\vartheta}= P_3^{\vartheta} (\xi) \, ,
\end{equation}
which has the solution
\begin{equation}
	\xi(\gamma) = \wp\left(\gamma - \gamma''_{\rm in}; g_2^{\vartheta}, g_3^{\vartheta}\right) \ ,
\end{equation}
where $\gamma''_{\rm in}=\gamma_{\rm in}+\int^\infty_{\xi_{\rm in}}{\frac{\dd \xi}{\sqrt{4\xi^3-g_2^{\vartheta}\xi-g_3^{\vartheta}}}}$ with $\xi_{\rm in}=\pm\frac{b'_3}{4\nu_{\rm in}} + \frac{b'_2}{12}$. Here the constants are 
\begin{equation}
	g_2^{\vartheta}=\frac{{b'_2}^2}{12} - \frac{b'_1b'_3}{4} \, , \qquad  g_3^{\vartheta}=\frac{b'_1b'_2b'_3}{48} - \frac{b'_0{b'_3}^2}{16}-\frac{{b'_2}^3}{216} \ .
\end{equation}
Now we can state the solution of (\ref{eqn:theta-equation}) after resubstitution as
\begin{equation}
 \vartheta(\gamma) = \arccos{\left(\frac{b'_3}{4\wp\left(\gamma - \gamma''_{\rm in}; g_2^{\vartheta}, g_3^{\vartheta}\right)-\frac{b'_2}{3} }+\nu_Y\right)}
\end{equation}
\subsubsection{Hyperelliptic case}
To derive the full solution of eq. (\ref{eqn:theta-equation}) we substitute $\cos{\vartheta}=\pm\frac{1}{\nu}+\nu_{Y}$ and get
\begin{equation}
  \left(\nu\frac{\dd \nu}{\dd \gamma}\right)^2=\sum_{i=0}^5 b'_i\nu^i=:P_5^{\vartheta}(\nu) \, ,
\end{equation}
whose solution can be formulated as
\begin{equation}
\vartheta = \arccos{\left(\mp \frac{\sigma_2(\vec{\gamma'}_{\infty})}{\sigma_1(\vec{\gamma'}_{\infty})}+ \nu_Y\right)} \, ,
\end{equation}
with
\begin{equation}
 \vec{\gamma'}_{\infty} = \left(-\int_{\nu}^{\infty}{\frac{\dd \nu}{\sqrt{P_5^{\vartheta}(\nu)}}}, \gamma - \gamma_{\rm in}- \int_{\nu_{in}}^{\infty}{{\frac{\nu \dd \nu}{\sqrt{P_5^{\vartheta}(\nu)}}}}\right)^T
\end{equation}
\subsection{The $\phi$-equation}
We need to use the equations (\ref{eqn:r-equation}) and (\ref{eqn:theta-equation}) to rewrite the $\phi$-equation (\ref{eqn:phi-equation}) in the form
\begin{equation}
 \dd \phi =\frac{\ta\tXi(\tB E-\ta\tL\tXi)}{\tR}\frac{\dd \tr}{\sqrt{X}}+\frac{\tXi(\tA E-\tL\tXi)}{\tTh\sin{\vartheta}}\frac{\dd \vartheta}{\sqrt{Y}}
\end{equation}
and in integral form
\begin{equation}
 \phi -\phi_{in} = \int_{\tr_{\rm in}}^{\tr}{\frac{\ta\tXi(\tB E-\ta\tL\tXi)}{\tR}\frac{\dd \tr}{\sqrt{X}}} + \int_{\vartheta_{\rm in}}^{\vartheta}{\frac{\tXi(\tA E-\tL\tXi)}{\tTh\sin{\vartheta}}\frac{\dd \vartheta}{\sqrt{Y}}} = I_{\tr}(\tr)+I_{\vartheta}(\vartheta) \, .
\end{equation}
$I_{\tr}$ and $I_{\vartheta}$ can be solved separately.
\subsubsection{Elliptic case}
Again in the case of $\tg=0$ or $\delta=0$ the solution is much simpler. We start with the $\tr$-dependant integral $I_{\tr}$ and substitute $\tr= \pm\frac{b_3}{4y-\frac{b_2}{3}}+\tr_{X}$ and apply a partial fraction decomposition
\begin{equation}
 I_{\tr}= \int_{y_{\rm in}}^{y} \left(C_0+ \sum_{i=1}^2 \frac{C_i}{y'-p_i}\right)\frac{\dd y'}{\sqrt{P^{\tr}_3(y')}} \, ,
\end{equation}
where $C_i$ are constants which arise from the partial fraction decomposition and $p_i$ are first order poles of $I_{\tr}$. 
With the last substitution $y=\wp(v';g_2^{\tr},g_3^{\tr})=\wp_{\tr}(v')$ , where $v'=\gamma-\gamma'_{\rm in}$, we get
\begin{equation}
 I_{\tr}= \int_{v'_{\rm in}}^{v'} \left(C_0+ \sum_{i=1}^2 \frac{C_i}{\wp_{\tr}(v')-p_i}\right)\dd v' \, ,
\end{equation}
The $\vartheta$-dependent integral $I_{\vartheta}$ can be stated in a similar form with the substitution \\
 $\cos{\vartheta}=\pm \frac{b'_3}{4{\wp_{\vartheta}(v'')}-\frac{b'_2}{3}}+\nu_y$
\begin{equation}
 I_{\vartheta}= \int_{v''_{in}}^{v'}{\left(C'_0+\sum_{i=1}^2\frac{C'_i}{\wp_{\vartheta}(v'')-p'_i}\right) \dd v''} \, ,
\end{equation}
where $C'_i$ are the corresponding constants from the partial fraction decomposition, $p'_i$ are the first order poles of $I_{\vartheta}$ and $\wp_{\vartheta}(v'')=\wp(v'';g_2^{\vartheta},g_3^{\vartheta})$, with $v''=\gamma-\gamma''_{\rm in}$. Both integrals are third kind elliptic integrals and can be solved in terms of the Weierstra{\ss} $\wp$, $\sigma$ and $\zeta$ functions (see \cite{Kagramanova:2010bk, Grunau:2010gd, Enolski:2011id}). Therefore we can state the solution of eq. (\ref{eqn:phi-equation}) as
\begin{equation}
\begin{split}
 \phi(\gamma) -\phi_{\rm in}=&C_0(v'-v'_{\rm in}) + \sum_{i=1}^2\left(2\zeta_{\tr}(v'_i)(v'-v'_{\rm in})+\ln{\frac{\sigma_{\tr}(v'-v'_i)}{\sigma_{\tr}(v'_{\rm in}-v'_i)}}-\ln{\frac{\sigma_{\tr}(v'+v'_i)}{\sigma_{\tr}(v'_{\rm in}+v'_i)}}\right) \\
	      &C'_0(v''-v''_{\rm in}) + \sum_{i=1}^2\left(2\zeta_{\vartheta}(v''_i)(v''-v''_{\rm in})+\ln{\frac{\sigma_{\vartheta}(v''-v''_i)}{\sigma_{\vartheta}(v''_{\rm in}-v''_i)}}-\ln{\frac{\sigma_{\vartheta}(v''+v''_i)}{\sigma_{\vartheta}(v''_{\rm in}+v''_i)}}\right) \, ,
\end{split}
\label{eqn:phi-solution1}
\end{equation}
whith $p_i=\wp_{\tr}(v'_i)$, $p_i=\wp_{\vartheta}(v''_i)$ and 
\begin{eqnarray}
	\wp_{\tr}(v) &= \wp (v, g_2^{\tr}, g_3^{\tr})\, , \qquad \wp_\vartheta (v)&= \wp (v, g_2^{\vartheta}, g_3^{\vartheta}) \, ,\nonumber\\
	\zeta_{\tr}(v) &= \zeta (v, g_2^{\tr}, g_3^{\tr})\, , \qquad \zeta_\vartheta (v)&= \zeta (v, g_2^{\vartheta}, g_3^{\vartheta}) \, ,\\
	\sigma_{\tr}(v) &= \sigma (v, g_2^{\tr}, g_3^{\tr})\, , \qquad \sigma_\vartheta (v)&= \sigma (v, g_2^{\vartheta}, g_3^{\vartheta}) \, .\nonumber
\end{eqnarray}
\subsubsection{Hyperelliptic case}
For the full solution we can write the integrals $I_{\tr}$ and $I_{\vartheta}$ in a form like
\begin{equation}
 I= \int_{x_{\rm in}}^{x} \frac{\dd x'}{(x-Z)\sqrt{P_5(x)}} \, ,
\end{equation}
where $P_5(x)$ is a polynomial of fifth order in $x$ and Z is pole. This is a hyperelliptic integral of the third kind and following \cite{Enolski:2011id} it can be solved by
\begin{equation}
 I=\frac{2}{\sqrt{P_5{(Z)}}}\int_{x_{\rm in}}^x{\dd\vec{z}^T}\int_{e_2}^{Z}{\dd\vec{y}}
+\ln{\left(\frac{\sigma(\int_{\infty}^x{\dd\vec{z}}-\int_{e_2}^Z{\dd\vec{z}})}
{\sigma(\int_{\infty}^x{\dd\vec{z}}+\int_{e_2}^Z{\dd\vec{z}})}\right)}
-\ln{\left(\frac{\sigma(\int_{\infty}^{x_{\rm in}}{\dd\vec{z}}-\int_{e_2}^Z{\dd\vec{z}})}
{\sigma(\int_{\infty}^{x_{\rm in}}{\dd\vec{z}}+\int_{e_2}^Z{\dd\vec{z}})}\right)} \, ,
\label{eqn:phi-solution2}
\end{equation}
where $e_2$ is a zero of $P_5(x)$ and
\begin{equation}
 \dd\vec{z}:=\left(\frac{\dd x}{\sqrt{P_5(x)}},\frac{x\dd x}{\sqrt{P_5(x)}}\right)^T
\end{equation}
\begin{equation}
 \dd\vec{y}=\left(\sum_{k=1}^4k a_{k+1}\frac{x^k\dd x}{4\sqrt{P_5(x)}},
\sum_{k=2}^3(k-1)a_{k+3}\frac{x^k\dd x}{4\sqrt{P_5(x)}}\right)^T
\end{equation}
are holomorphic and meromorphic differentials. Here $a_k$ denote the coefficients of the polynomial $P_5(x)$.
\section{The $\tlt$-equation}
The $\tlt$-equation (\ref{eqn:t-equation}) can be solved analogously to the $\phi$-equation (\ref{eqn:phi-equation}) and has solutions of the form of eq. (\ref{eqn:phi-solution1}) for $\tg=0$ or $\delta=0$ or of the form of eq. (\ref{eqn:phi-solution2}) for the full solution. 

\section{The orbits}
\begin{figure}[p]
	\centering
	\subfigure[$\delta=1$, $\ta=-0.4$, $\tg=0.2$, $\tN=-0.07$, $\te=0.07$, $\tv=1.2$, $\tK=-0.1$, $\tL=-1$, $E=2.8$: Bound orbit]{
		\includegraphics[width=0.31\textwidth]{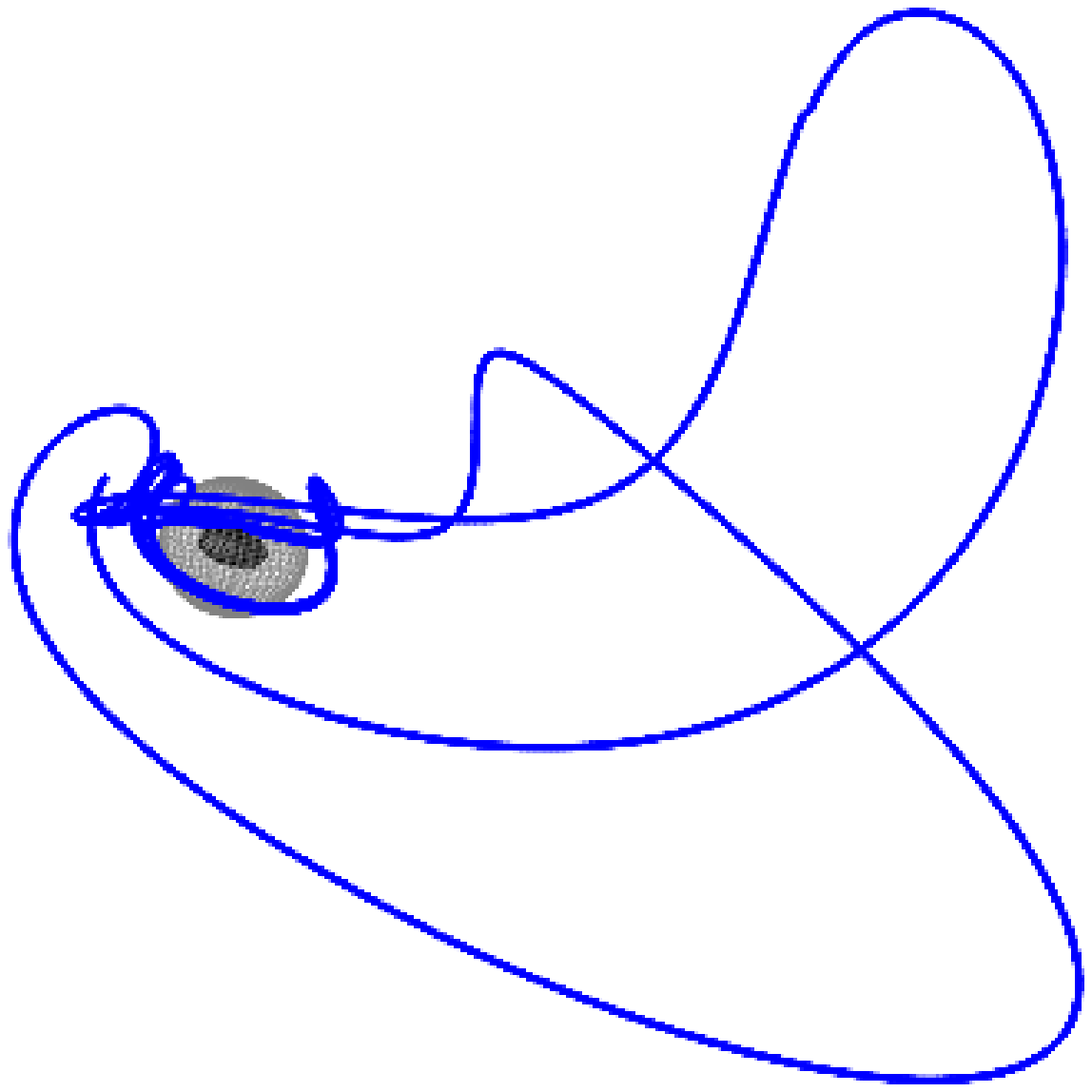}
	}
	\subfigure[$\delta=1$, $\ta=0.4$, $\tg=0$, $\tN=0$, $\te=0$, $\tv=0$, $\tK=-10$, $\tL=2.3$, $E=1.953$: Escape Orbit]{
		\includegraphics[width=0.31\textwidth]{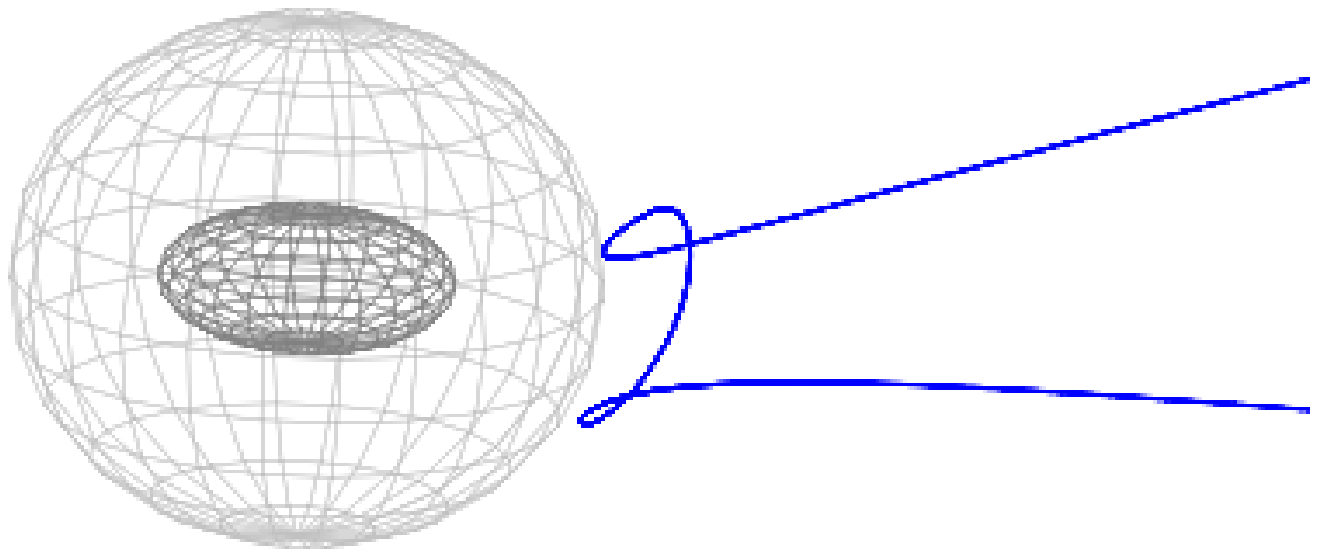}
	}
	\subfigure[$\delta=1$, $\ta=0.4$, $\tg=0$, $\tN=0$, $\te=0$, $\tv=0$, $\tK=-10$, $\tL=2.3$, $E=1.953$: Many-world bound orbit]{
		\includegraphics[width=0.31\textwidth]{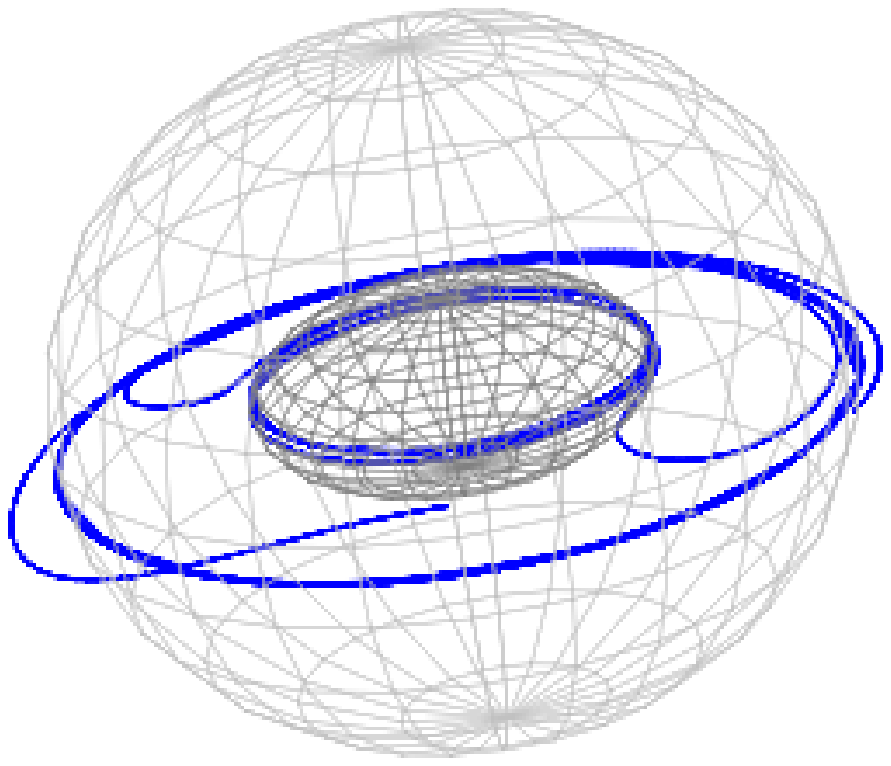}
	}
	\caption{Orbits of test particles in the dyonic rotating black hole spacetime. The orbit is represented by the blue line and the outer and inner horizon are denoted by grey and black ellipsoids.}
 \label{pic:orbits-1}
\end{figure}

\begin{figure}[p]
	\centering
	\subfigure[$\delta=1$, $\ta=-0.4$, $\tg=0$, $\tN=0$, $\te=0$, $\tv=0.2$, $\tK=-1$, $\tL=4.6$, $E=12$: Bound orbit]{
		\includegraphics[width=0.31\textwidth]{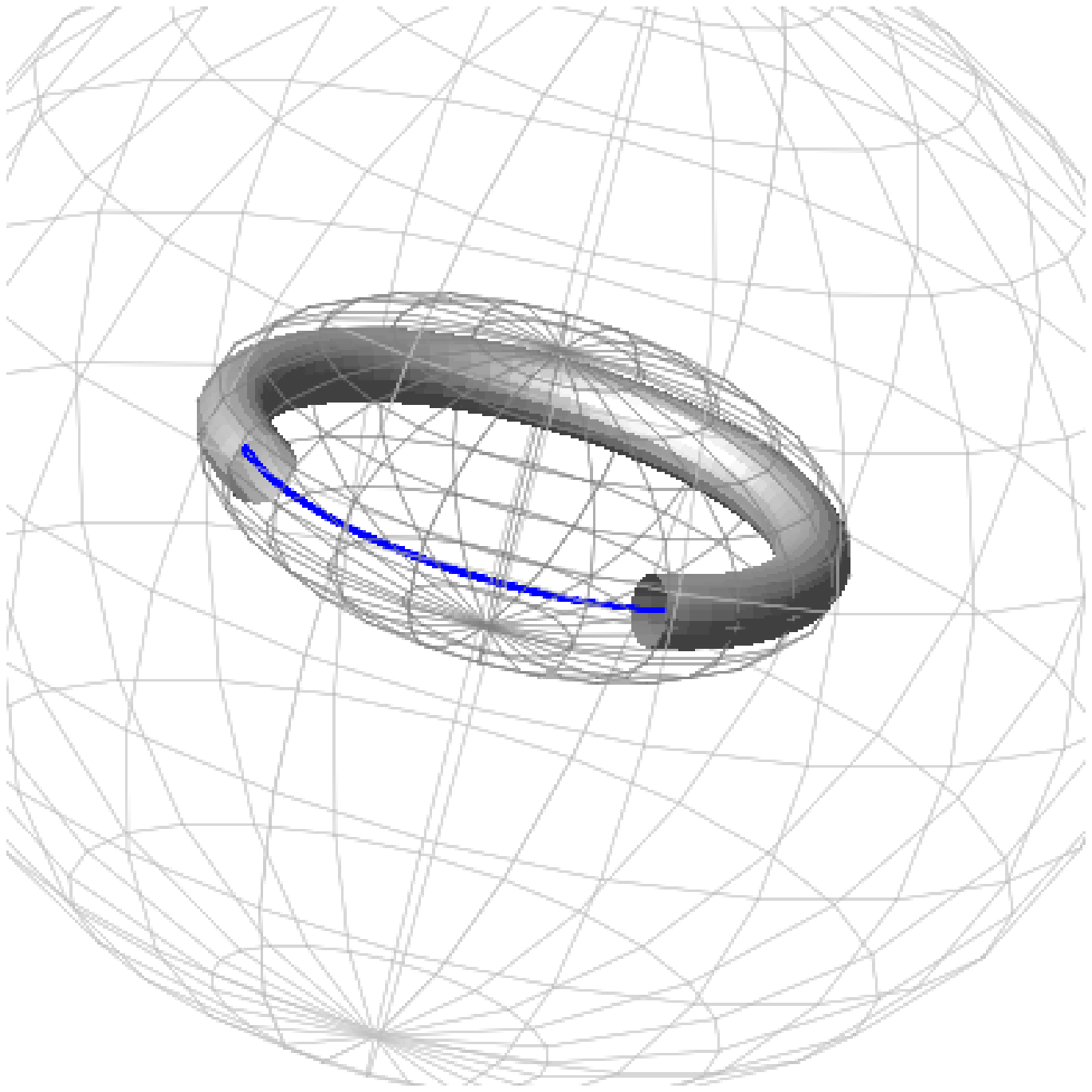}
	}
	\subfigure[$\delta=1$, $\ta=0.3$, $\tg=0$, $\tN=0.2$, $\te=0.2$, $\tv=0.1$, $\tK=-0.1$, $\tL=3$, $E=4$: Escape orbit with $\tr<0$]{
		\includegraphics[width=0.31\textwidth]{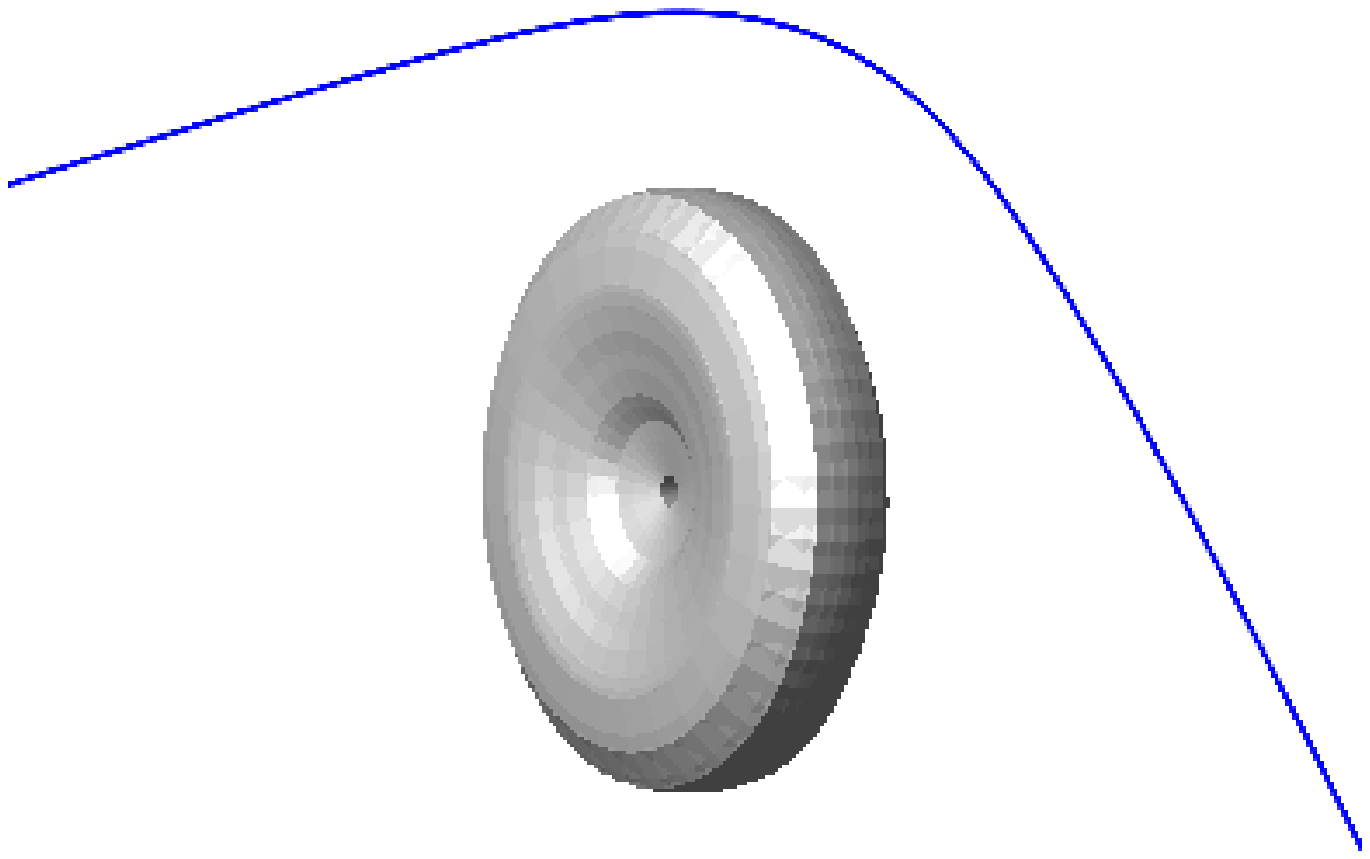}
	}
	\subfigure[$\delta=1$, $\ta=0.4$, $\tg=0$, $\tN=0.35$, $\te=0.35$, $\tv=0.1$, $\tK=0.1$, $\tL=3$, $E=2$: Two-world escape orbit]{
		\includegraphics[width=0.31\textwidth]{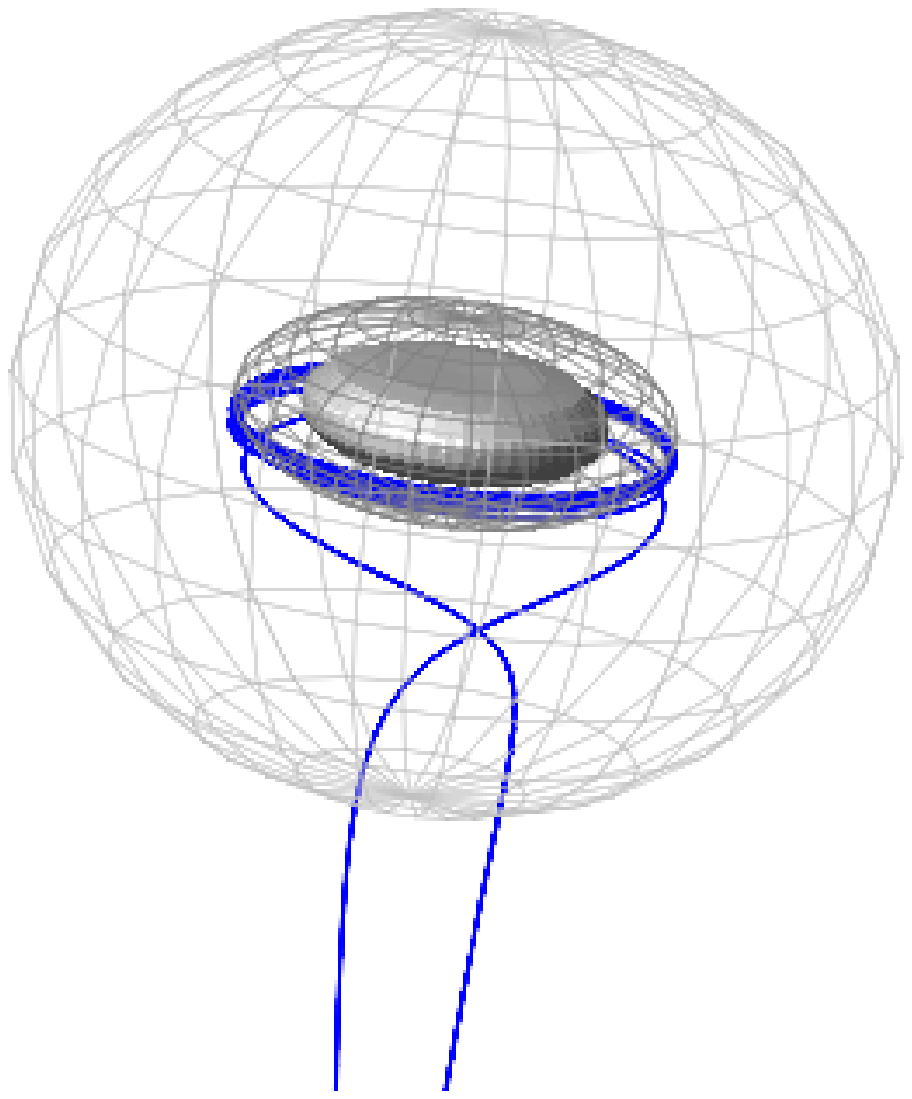}
	}
	\caption{Orbits of test particles in the dyonic rotating black hole spacetime. The orbit is represented by the blue line and the outer and inner horizon are denoted by grey and black ellipsoids. The grey structure denotes the singularity.}
 \label{pic:orbits-2}
\end{figure}

Now we can use the analytical solutions to visualize the possible orbits in this spacetime. In fig. \ref{pic:orbits-1} examples of a bound orbit, an escape orbit and a many-world bound orbit are shown.  Fig. \ref{pic:orbits-1}(a) shows the special case of a zoom-whirl orbit, where the particle makes several turns near the horizon, then escapes to a region far away of the black hole and turns back to the near horizon regime. Figure \ref{pic:orbits-2}(a) shows a BO which lies inside the singularity. We cut out a part of the ringoid structure to make it visible. The orbit in fig \ref{pic:orbits-2}(b) shows an escape orbit around a naked singularity for negative $\tr$. The singularity has a toroidal form and it may be possible to fly through the hole in the middle. The last orbit in fig \ref{pic:orbits-2}(c) shows a two-world escape orbit, where the turning point is very close to the inner horizon. The singularity for this black hole is an ellipsoid.  

\section{Conclusion}
In this paper we derived the equations of motion for test particles and light in the $\text{U(1)}^2$ dyonic rotating black hole spacetime and stated their analytical solution in two cases. For the full solution we had to use methods for solving hyperelliptic differential equations of the first and third kind. For vanishing cosmological constant or light the solution can be formulated in an easier form, since we only had to solve elliptic differential equations. We were able to characterize the possible orbit types for two-horizon black holes. These were orbits we already know from the special cases of the spacetime like escape orbits and bound orbits. A special feature are bound orbits and many-world bound orbits, which crosses $\tr=0$. We call these CBO or CMBO. In addition we found an interesting structure of the singularity varying from toruslike structures to ellipsoids for $\tr \geq 0$ in contrast to the EMDA black hole $\cite{Flathmann:2015}$. 

Future work could be the characterization of geodesic motion for black holes with more than two horizons, especially the influence of the virtual horizon on the particle motion. Another step is to derive the equations of motion for charged particles or to consider an even more general black hole with more charges. In these cases the differential equation we have to deal with are at least hyperelliptic of higher genus or even only solvable in special cases. The analytical solutions we derived can now be used to calculate observables like the periastron shift, the light deflection or the shadow of the black hole, which eventually can be compared to observations in the future.   

\section{Acknowledgements}
We would like to thank Jutta Kunz for helpful and interesting discussions. S. G. gratefully acknowledges support by the DFG, within the Research Training Group \textit{Models of Gravity}.

\bibliographystyle{unsrt}

\end{document}